\documentclass[journal]{IEEEtran}
\ifCLASSINFOpdf
\else
\fi
\usepackage[utf8]{inputenc}
\usepackage{amssymb}
\usepackage{tcolorbox}
\tcbuselibrary{breakable}
\usepackage{bbm}
\usepackage{amsfonts}
\PassOptionsToPackage{dvips}{graphicx}
\usepackage{times}
\usepackage[]{cite}
\usepackage{amsmath}
\usepackage{array}
\usepackage{amssymb}

\usepackage{stfloats}
\usepackage{slashbox}
\usepackage{footnote}
\usepackage{booktabs}
\usepackage{array}
\usepackage{mathtools}
\usepackage{dsfont}
\usepackage{ntheorem}
\usepackage{subeqnarray}
\usepackage{cases}
\usepackage{threeparttable}
\usepackage{color}
\usepackage{nomencl}
\makenomenclature
\usepackage[numbers,sort&compress]{natbib}
\IEEEoverridecommandlockouts

\newtheorem{remark}{Remark}

\usepackage{flafter}
\usepackage{subfigure}
\usepackage{float}
\usepackage{bbold}
\usepackage{supertabular}
\usepackage{caption}
\usepackage{url}
\usepackage{xcolor}
\usepackage[ruled,vlined,lined,ruled,linesnumbered]{algorithm2e}
\newcommand{\bm}[1]{\mbox{\boldmath{$#1$}}}

\usepackage{subfigure}
\usepackage{stfloats}
\usepackage{etoolbox}
\renewcommand\nomgroup[1]{%
  \item[\bfseries
  \ifstrequal{#1}{P}{Physics Constants}{%
  \ifstrequal{#1}{N}{Number Sets}{%
  \ifstrequal{#1}{O}{Other Symbols}{}}}%
]}
\allowdisplaybreaks

\usepackage[textsize=tiny]{todonotes}

\begin{document}
%
\title{\huge{Universal Graph Learning for Power System Reconfigurations: Transfer Across Topology Variations}}
\IEEEaftertitletext{\vspace{-2.2\baselineskip}}
%

\author{Tong Wu, \IEEEmembership{IEEE Member}, Anna Scaglione, \IEEEmembership{IEEE Fellow}, Sandy Miguel,  Daniel Arnold, \IEEEmembership{IEEE Member}
\thanks{Tong Wu is with the Department of Electrical and Computer Engineering, University of Central Florida, Orlando, FL, 32816 USA (e-mail:   tong.wu@ucf.edu).   Anna Scaglione   is with the Department of Electrical and Computer Engineering, Cornell Tech, Cornell University, New York City, NY, 10044 USA (e-mail:  as337@cornell.edu). Daniel Arnold is with Lawrence Berkeley National Laboratory (e-mail: dbarnold@lbl.gov).  }

\thanks{This work was supported in part by National Science
Foundation (NSF) under Grant NSF ECCS 2210012, and in part by the Cybersecurity, Energy Security, and Emergency Response (CESER),  Risk Management
and Tools and Technologies (RMT) Program of the U.S. Department of Energy
through the "Mitigation via Analytics for Inverter-Grid Cybersecurity (MAGIC)" Project under contract DE-AC02-05CH11231.}
 }

\newcommand{\norm}[1]{\left\lVert#1\right\rVert}
\newcommand*\abs[1]{\lvert#1\rvert}


\maketitle

\begin{abstract}
This work addresses a fundamental challenge in applying deep learning to power systems: developing neural network models that transfer across significant system changes, including networks with entirely different topologies and dimensionalities, without requiring training data from unseen reconfigurations. Despite extensive research, most ML-based approaches remain system-specific, limiting real-world deployment. This limitation stems from a dual barrier. First, topology changes shift feature distributions and alter input dimensions due to power flow physics. Second, reconfigurations redefine output semantics and dimensionality, requiring models to handle configuration-specific outputs while maintaining transferable feature extraction.
To overcome this challenge, we introduce a Universal Graph Convolutional Network (UGCN) that achieves transferability to any reconfiguration or variation of existing power systems without any prior knowledge of new grid topologies or retraining during implementation. Our approach applies to both transmission and distribution networks and demonstrates generalization capability to completely unseen system reconfigurations, such as network restructuring and major grid expansions. Experimental results across power system applications, including false data injection detection and state forecasting, show that UGCN significantly outperforms state-of-the-art methods in cross-system zero-shot transferability of new reconfigurations.
\end{abstract}


\section{Introduction}
\subsection{Background and Motivation}
Traditional optimization and control methods in power systems, such as model-based optimization and rule-based control, depend on accurate system models and well-defined mathematical formulations \cite{zhu2015optimization}. However, the growing complexity, uncertainty, and real-time operational demands of modern power grids—especially with large-scale distributed energy resources (DERs) and flexible demand control at the grid edge—challenge the scalability and adaptability of these approaches \cite{wu2023constrained}. Grid-edge systems, including smart buildings and electrified transportation, further exacerbate these challenges \cite{aguero2017modernizing}. AI-driven methods, particularly ML, provide a data-driven alternative that infers system dynamics, improves decision-making under uncertainty, and enables real-time control without the heavy cost of iterative optimization by leveraging historical data \cite{su2024review}. This allows power systems to operate with greater efficiency, resilience, and adaptability, overcoming the limits of traditional techniques \cite{wu2023complex}.

Although supervised ML approaches, particularly neural networks (NNs), have shown great potential in power systems, a critical limitation remains: most models fail to generalize beyond the specific systems on which they are trained \cite{li2022transfer, wu2024transferable}. Traditional NN approaches rely on data from specific power networks, making them sensitive to topology, operating conditions, and measurement availability \cite{khodayar2020deep}. As a result, they often fail or become unusable under major topology or parameter changes, limiting scalability and practical use \cite{hijazi2023transfer}.
This lack of transferability creates multiple deployment challenges. First, retraining for each new system, makes industrial deployment prohibitively expensive and impractical for real-time applications \cite{niu2021decade}. Second, operators cannot easily detect when NN models become outdated, rendering NN-based systems untrustworthy. Additionally, this inability to adapt limits the effectiveness of NN approaches when applied to rapidly evolving grid conditions \cite{wu2020voltage}. 

To overcome these challenges, we develop a Universal Graph Convolutional Network (UGCN) that achieves transferability to any reconfiguration of existing power systems without prior knowledge of new grid topologies or retraining during implementation.



\subsection{Related Work}
Recent advancements in ML/NN for power systems have primarily focused on three approaches to enhance model transferability: \textit{domain adaptation}, \textit{fine-tuning approaches}, and \textit{feature-based transfer learning}.
\textbf{Domain adaptation} techniques~\cite{shi2022power, wu2020voltage} attempt to align feature distributions between source and target power systems. However, these methods falter on heterogeneous grids with varying graph structures. They assume similar feature distributions across networks—an assumption violated when topology changes alter power flow physics. Additionally, these approaches still require target-domain labeled data and perform poorly with significant domain gaps, as illustrated when electrical islands merge or split.
\textbf{Fine-tuning approaches}~\cite{li2022transfer, li2022adaptive, ren2021integrated} pre-train models on source systems and then adapt them to target systems through additional training. While reducing initial training costs, these methods fundamentally require labeled data from each new configuration and assume identical label spaces between domains. This is impractical for real-time grid operations, where reconfigurations occur frequently, as system operators cannot collect and label data for every topology change. Meta-learning variants~\cite{zheng2024meta, sun2019meta, xia2024efficient} fail similarly, as they still need labeled examples from unseen configurations.
\textbf{Feature-based transfer learning}~\cite{kim2023transient, shi2022bidirectional, liang2024unified} attempts to extract invariant representations across power systems. For instance, \cite{liang2024unified} introduced plug-and-play neurons to accommodate growing networks. However, their method remains essentially system-to-system transfer learning rather than a universal framework for power system classes. It supports network expansion but not arbitrary reconfigurations, and when scaling to larger systems it suffers from parameter overwriting, causing knowledge learned from smaller networks to be lost and hindering the development of a truly universal model. 
Existing approaches share key flaws: they rely on labeled data that are system specific, assume aligned features or labels, break down on heterogeneous topologies, and perform poorly under large domain gaps.

\subsection{Essential Challenges}
The limitations of existing transfer learning approaches stem from fundamental challenges intrinsic to power system reconfiguration: their input and output distribution and dimensionality differences.

\subsubsection{Input Feature Distribution and Dimensionality Difference}

Power system reconfiguration prevents the use of conventional ML models by fundamentally altering both feature distributions and dimensionality. Each topology change---whether a single line break or bus reconfiguration---shifts the entire feature space due to power flow physics, creating partially overlapping or entirely distinct   statistical distributions that invalidate trained models. In more severe cases, entire distributions merge when system reconfigurations cause previously separate electrical islands to interconnect, making the feature space unrecognizable to the original model. 

Beyond distribution shifts, the input dimensionality itself becomes variable: removing a line reduces observable states, while adding buses increases them. This is not a simple scaling problem---it represents a fundamental incompatibility where models trained on $n$-dimensional inputs cannot process $(n\pm k)$-dimensional data without complete retraining. Even advanced graph neural networks, designed to handle varying graph sizes, fail here because they cannot simultaneously accommodate the dimensional mismatch and the physics-induced distribution shifts that render features from different topologies statistically incomparable.

\subsubsection{Output Label Distribution and Dimensionality Difference}

The output space presents an even more fundamental challenge that exposes the limitations of existing transfer learning paradigms. System reconfiguration does not merely shift output distributions---it fundamentally redefines the semantic meaning of outputs. For example, a false data injection attack detected at bus 5 in one configuration has no meaningful correspondence to bus 5 in a reconfigured system where connection patterns have changed. When systems expand from $m$ to $m+p$ buses, models must suddenly predict $p$ additional state variables without historical training data on them, while the label distribution of the original $m$ buses also shifts; conversely, during outages, the model must handle dimension reduction while maintaining physical consistency.

This reveals a critical design paradox for achieving universal transferability. While successful transfer learning typically relies on finding invariant features in the input space that generalize across domains, power system outputs are inherently configuration-dependent and cannot share representations. A truly universal model faces conflicting requirements: it must extract transferable patterns from inputs while simultaneously adapting to completely independent output distributions for each configuration. Traditional approaches that update parameters for new topologies inevitably overwrite previous knowledge, preventing true universality. The model must instead maintain the capacity to generalize to any output label distribution without catastrophic forgetting---a challenge made particularly acute by physics constraints. Unlike domains such as computer vision where outputs can be interpolated or padded, power system states must strictly satisfy Kirchhoff's laws and power balance equations.


\subsection{Our Contribution}
We discuss next our technical contributions, which synergistically break critical barriers to achieve universal, zero-shot transferability across arbitrary power system reconfigurations. 
\subsubsection{Topology Evolution as a Learnable Dimension} We reconceptualize power network reconfiguration treating topology changes as an additional dimension alongside space and time. Instead of treating major network reconfigurations as completely different problems requiring new models, we treat them as points along the ``topology dimension." Through physics-aware augmentation and by training on multiple system configurations simultaneously, we create a higher-dimensional learning space where all topology variations—despite having different distributions and dimensions—coexist in one  dataset. This enables the model to recognize that different configurations share underlying power flow patterns and physical relationships, making it possible to transfer knowledge to completely new topologies without any retraining.

\subsubsection{Application-Oriented Adaptive Pooling with Scalar-Weight Graph Convolutional Neural Networks (GCN) for Input Invariance} To solve the dual barrier of feature distribution shifts and dimensional mismatches, we introduce adaptive pooling mechanisms paired with scalar-weight complex-valued GCN layers. The scalar weights—constant regardless of network size—enable feature extraction across networks of any dimension, while the complex-valued formulation preserves magnitude and phase information critical for power system state analysis. The novel insight is that while reconfiguration creates partially overlapping or entirely distinct distributions with varying dimensions, these distributions share common patterns when viewed through application-specific lenses. Our application-oriented adaptive pooling exploits this by extracting invariant features: maximum pooling captures high-frequency graph signals (anomaly signatures) that persist across any topology, while learnable pooling preserves low-frequency spatial correlations essential for state estimation. This approach bridges variable input dimensions to fixed hidden layers while extracting shared features from disparate distributions, turning dimensional incompatibility and distribution shifts into a single solvable problem through strategic information extraction rather than forced alignment.

\subsubsection{Parallel Transformer for Output Dimension-Distribution Adaptation} To address the paradox where models must maintain transferable input features while generating configuration-specific outputs with varying dimensions and distributions, we develop a parallel transformer architecture that processes each configuration's output space independently. The challenge is fundamental: output semantics change completely with reconfiguration, and dimensional variations require generating entirely new state variables without training data. Our solution employs network-specific positional encodings with parallel decoder branches, where each configuration maintains its own output space while sharing learned feature representations. This prevents catastrophic forgetting—where updating parameters for new topologies destroys previously learned knowledge. By preserving all configuration-specific knowledge simultaneously through parallel processing rather than sequential overwriting, the transformer dynamically generates outputs matching any target dimension while maintaining physical constraints, achieving zero-shot transferability to arbitrary reconfigurations.

\subsection{Applications}
Beyond these technical innovations, we highlight potential applications of the UGCN framework next.
\subsubsection{Distribution System Reconfigurations} Distribution networks undergo frequent structural changes, including multiple line failures, feeder additions or removals, and parameter variations. The UGCN enables a single model to adapt dynamically to these variations, eliminating the need for retraining while ensuring reliable performance in critical tasks such as power system state forecasting. We validate this through extensive testing on radically reconfigured distribution feeders, including topology changes that would typically require entirely new models.
\subsubsection{Multiple Heterogeneous Transmission Systems}
Another advantage of the UGCN is its ability to learn from multiple heterogeneous transmission systems simultaneously within one framework, adapting to reconfigurations across all trained systems without forgetting. For example, a single model could be trained to operate on both the California and Texas power grids—two systems with completely different designs, voltage levels, generation mixes, and operating characteristics, yet no direct connection to each other. Unlike traditional methods that need a separate model for each grid, our UGCN learns from multiple unrelated systems simultaneously, retaining performance under reconfigurations without forgetting. It operates seamlessly on trained grids and transfers to entirely new ones, as shown in our FDI detection across diverse transmission systems.

\begin{remark}
For a single large grid transferring to its reconfigurations, the UGCN handles this directly without dimensional challenges since reconfigurations typically maintain similar scales. However, for achieving cross-regional transferability between grids with vastly different dimensions, the framework would benefit from an extension using graph decomposition. Decomposing large networks into subgraphs would enable training on manageable subgraphs while maintaining global consistency— an extension we defer to future work.
\end{remark}

The remainder of the paper is structured as follows. Section II introduces the preliminaries of the physics-aware grid GCN and the grid-graph generation process for cross-graph training. Section III presents the universal GCN framework, highlighting three key novel techniques for transferability. In Section IV, we apply the proposed universal GCN to power system applications, including state forecasting and false data injection detection. Case studies are provided in Section V. Finally, we conclude the paper in Section VI.

\section{Preliminaries}
In this section, we first examine the design of physics-aware graph neural networks and explain why they are the best choice for transferability in power systems.

\subsection{Basic Notations}
Electric grid networks have an associated undirected weighted graph, denoted as $\mathcal{G}(\mathcal{V}, \mathcal{E})$. In this graph, the \textit{buses} correspond to the nodes, and the \textit{transmission lines} are represented as edges. The relationships between the current and voltage phasors for the network are described by Kirchhoff's and Ohm's laws, which can be written as follows:
\begin{equation}
\begin{aligned}
 \bm{i} = \mathbf{Y} \bm{v}, \quad  v_{n} = |v_{n}|~e^{\mathfrak{j} \varphi^v_{n}}, \quad 
     i_{n} =  |i_{n}|~e^{\mathfrak{j} \varphi^i_{n}}, \quad \forall {n} \in \mathcal{V}
 \label{eq:node_injection}
\end{aligned}
\end{equation}
where $\bm{v} \in \mathbb{C}^{N}$ and $|\bm{v}| \in \mathbb{R}_+^{N}$ represent the vectors of bus voltage phasors and magnitudes, respectively, and $\bm{i} \in \mathbb{C}^{N}$ and $|\bm{i}| \in \mathbb{R}_+^{N}$ are the vectors of net bus current phasors and magnitudes, respectively. The imaginary unit is $\mathfrak{j} = \sqrt{-1}$.

To apply graph signal processing (GSP) techniques to power systems, we establish the connection between physical power flow equations and graph-based representations. In GSP, the admittance matrix inverse $\mathbf{Y}^{-1}$ acts as a smoothing operator that propagates signals across network neighborhoods, behaving as a \textit{low-pass graph filter} \cite{wu2023complex}. Hence, Ohm's law suggests that we can interpret voltages as the output of a low-pass filtering operation: $\bm{v} = \mathbf{Y}^{-1} \bm{i}$.

\subsection{Physics-Aware Grid GCN}
To enhance the feature extraction capabilities compared to traditional neural networks, graph filters can be leveraged for improved representation learning. To define these filters, we first introduce key concepts, starting with the Graph Shift Operator (GSO).
A graph signal, denoted by $\bm{x} \in \mathbb{C}^{N}$, is a vector indexed by the nodes of a graph; for example, it represents the state vector of voltage phasors at each bus in a power grid. The neighborhood of node $i$, represented as $\mathcal{N}(i)$, refers to the set of nodes directly connected to node $i$.
The GSO, represented by the matrix $\mathbf{S} \in \mathbb{C}^{N \times N}$, is a neighborhood operator that combines values from neighboring nodes. We consider complex symmetric GSOs, i.e., $\mathbf{S} = \mathbf{S}^\top$. Designed to mimic a differential operator, the GSO is commonly selected as a graph weighted Laplacian matrix that incorporates the physical properties of the power system \cite{wu2023spatio}.

A graph filter is a linear matrix operator $\mathcal{H}(\mathbf{S})$, which is a function of the GSO. It acts on graph signals as follows:
\begin{align}\label{gs}
    \tilde{\bm{x}}^{(1)} = \mathcal{H}(\mathbf{S}) \bm{x}^{(0)}
\end{align}
where $\bm{x}^{(0)}$ denotes the input features. In power systems applications, $\bm{x}^{(0)}$ is typically the state vector $\bm{v}$.

The key characteristic of the graph filter $\mathcal{H}(\mathbf{S})$ is that it must be shift-invariant with the GSO, similar to time-invariant filters in the time domain. This means that $\mathcal{H}(\mathbf{S})$ must satisfy the condition $\mathcal{H}(\mathbf{S})\mathbf{S}=\mathbf{S}\mathcal{H}(\mathbf{S})$. This property holds only if $\mathcal{H}(\mathbf{S})$ can be expressed as a matrix polynomial:
\begin{equation}
\begin{split}\label{lsi}
 \mathcal{H}(\mathbf{S}) = \sum_{k=0}^{K} h_k  \mathbf{S}^k
\end{split}
\end{equation}
where the graph filter order $K$ can potentially be infinite. We can extend the spatial convolution to the spatio-temporal convolution version to capture both spatial and temporal information from voltage phasors:
\begin{equation}
\begin{split}\label{lsi2}
    \tilde{\bm{x}}^{(1)}_t   = \sum_{k=0}^{K} \sum_{\tau = 0 }^{K_t } h_{k, \tau}  \mathbf{S}^k \bm{x}^{(0)}_{t-\tau}
\end{split}
\end{equation}

Based on Eq.~\eqref{lsi2}, the GCN layer operation is:
\begin{equation}
    \bm{x}^{(\ell)}_t = \sigma \left[ \tilde{\bm{x}}^{(\ell)}_t \right] = \sigma\left[\sum_{k=0}^{K} \sum_{\tau = 0 }^{K_t } h_{k, \tau} \mathbf{S}^k \bm{x}^{(\ell-1)}_{t-\tau} \right]
\end{equation}
where $\ell = 1, \ldots, L$ represents the layer index. Here, $\bm{x}^{(0)} = \bm{v} \in \mathbb{C}^N$, $\mathbf{S}^k \in \mathbb{C}^{N \times N}$, and $h_{k,\tau} \in \mathbb{C}$ are learnable parameters. The activation function $\sigma(\cdot)$ is typically ReLU.

For the output layer, fully connected networks (FNNs) are often used, i.e. 
\begin{equation}  
    \bm{y} = \sigma(\mathbf{W} \bm{x}^{(L)})
\end{equation}  

GCNs exhibit superior transferability across graphs due to their parameter-sharing architecture and local processing characteristics \cite{wu2023complex}, stemming from scale-invariant parameters, where graph filter coefficients $h_{k, \tau} \in \mathbb{C}$ are scalars \textbf{independent of network size} $N$, enabling direct application across systems of different scales without modification. Additionally, graph convolutions operate through $k$-hop local aggregation, mirroring how electrical phenomena propagate in power systems, ensuring that learned features represent local power system behaviors. By incorporating admittance matrix properties into the GSO design, GCNs align with power flow physics, learning universal electrical relationships rather than system-specific patterns. However, redefining the training so that different systems can be sampled in learning process is of paramount importance and this is what distinguishes conventional GCN from the UGCN design we propose.

\begin{figure*}[!htp]
 \vspace{-0.2cm}
  \center
    \includegraphics[width=0.9\textwidth]{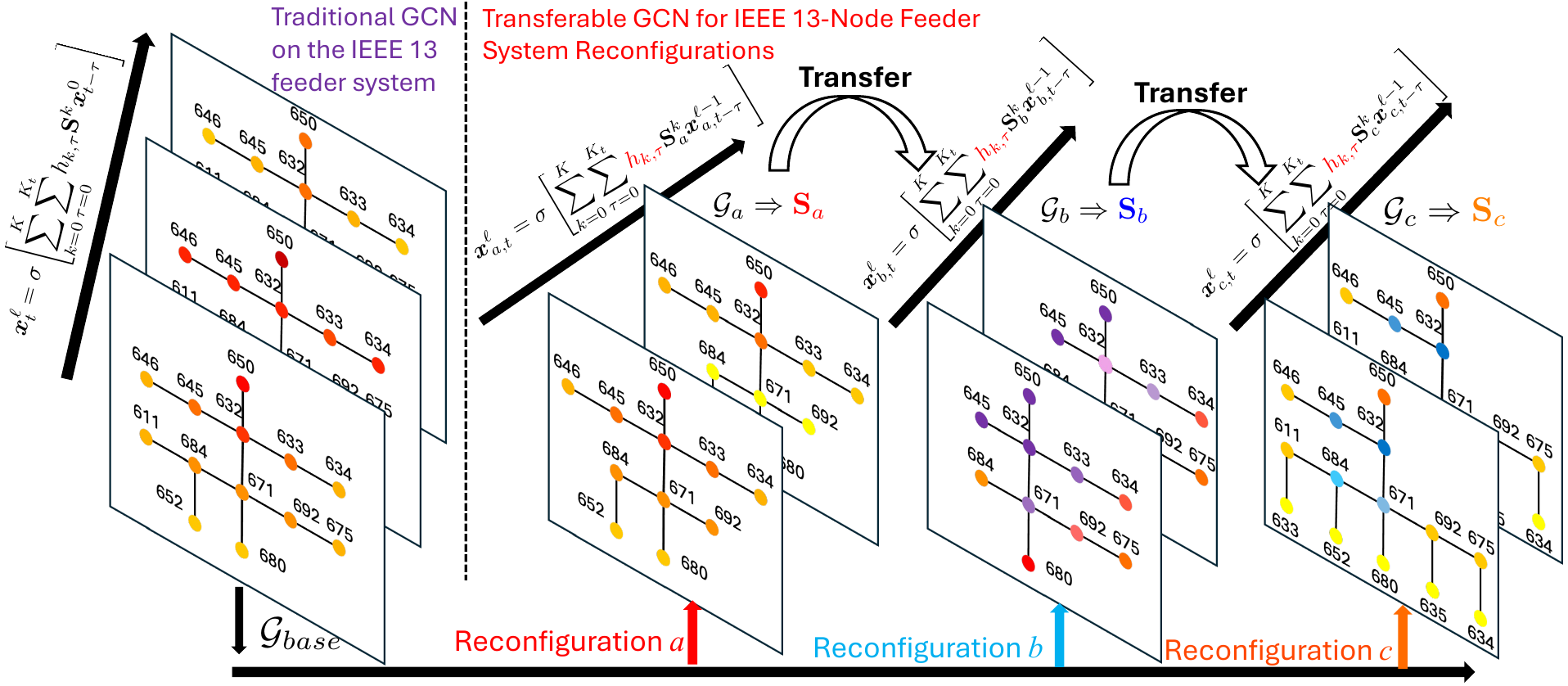}
  \caption{Comparison of Traditional GCN and Transferable UGCN in Distribution Systems Under Different Reconfigurations.
In the left figure, the trainable parameter \( h_{k, \tau} \) in the traditional GCN is trained specifically for a given topology. In contrast, in the right figure, the trainable parameter \( h_{k, \tau} \) remains unique across different grid reconfigurations, demonstrating universal transferability. }\label{transfer_gcn2}
  \vspace{-0.6cm}
\end{figure*}

\section{Universal Graph Convolutional Neural Network for Grid Reconfiguration}
To ensure that UGCNs remain transferable across different power systems $q \in \mathcal{Q}$ with varying graph shift operators $\mathbf{S}_q$ and input features $\bm{x}_q$, the key challenge lies in augmenting the training data and graph distributions to enable UGCN parameters to learn sufficiently broad representations. 

The first step involves enhancing training through physics-aware graph augmentation by systematically generating diverse network configurations and their corresponding operational states. This approach enables UGCNs to generalize to arbitrary grid reconfigurations, including topological changes not encountered during training. 

While GCN layers provide input dimension independence through their scale-invariant filter coefficients, a critical architectural challenge emerges in the transition from GCN layers to output layers. Specifically, we must maintain consistent hidden layer dimensions across networks with varying sizes while preserving the model's ability to generate outputs of different dimensions for different target systems. This dual requirement—universal feature extraction combined with adaptive output generation—forms the core technical contribution of our UGCN framework.

\subsection{Training Graph Augmentation}\label{SecIIIA}
To ensure the UGCN framework generalizes across different grid topologies, we propose a systematic approach to generate diverse training configurations from a single base power system. Starting with $\mathcal{G}_{\text{base}} = (\mathcal{V}_{\text{base}}, \mathcal{E}_{\text{base}})$, we apply various modifications to create multiple augmented variants $\{\mathcal{G}_q\}_{q=1}^Q$ that simulate realistic operational scenarios, where $Q$ is the total number of augmented systems and $q \in \mathcal{Q} = \{1, 2, \ldots, Q\}$. A distribution system can be represented as a tree, where $\mathcal{V}_q$ is the set of nodes (buses, substations, and distributed energy resources) and $\mathcal{E}_q$ represents edges (power lines and transformers) for system $q$. The root node corresponds to the substation supplying power to the distribution network, and the tree constraint ensures connectivity without cycles.

Distribution networks frequently undergo topology modifications due to operational changes, fault management, and grid expansion. The following key reconfiguration scenarios are applied to $\mathcal{G}_{\text{base}}$ to generate the training set $\{\mathcal{G}_q = (\mathcal{V}_q, \mathcal{E}_q)\}_{q=1}^Q$:
\begin{itemize}
\item \textbf{Feeder Disconnection:} A node $n \in \mathcal{V}_{\text{base}}$ and its associated edges are removed, while the reconfigured network retains a tree structure: 
$\mathcal{V}_q = \mathcal{V}_{\text{base}} \setminus \{ n \}$, $\mathcal{E}_q = \mathcal{E}_{\text{base}} \setminus \{ (n, m) \mid m \in \mathcal{V}_{\text{base}} \}$.

\item \textbf{New Feeder Connection:} A new node $n_{\text{new}}$ is added and connected to an existing node $p \in \mathcal{V}_{\text{base}}$, 
$\mathcal{V}_q = \mathcal{V}_{\text{base}} \cup \{ n_{\text{new}} \}$, $\mathcal{E}_q = \mathcal{E}_{\text{base}} \cup \{ (n_{\text{new}}, p) \}$.

\item \textbf{Parameter Changes:} The impedance of an edge $(n, m) \in \mathcal{E}_{\text{base}}$ is modified due to factors such as conductor aging, frequency or temperature variations, tap changing operations etc,: 
$Z'_{nm} = Z_{nm} + \Delta Z_{nm}$, where $Z_{nm}$ represents the impedance between nodes $n$ and $m$.

\item \textbf{Line Breaks:} If an edge $(n, m) \in \mathcal{E}_{\text{base}}$ is removed, leading to the disconnection of a subgraph, the updated edge set is: 
$\mathcal{E}_q = \mathcal{E}_{\text{base}} \setminus \{ (n, m) \}$.
If this results in a disconnected subtree $(\mathcal{V}_{\text{sub}}, \mathcal{E}_{\text{sub}})$ that does not contain the root node, the entire subtree is removed:
$\mathcal{V}_q = \mathcal{V}_{\text{base}} \setminus \mathcal{V}_{\text{sub}}$, $\mathcal{E}_q = \mathcal{E}_{\text{base}} \setminus \mathcal{E}_{\text{sub}}$.

\item \textbf{Subtree Merging:} A previously disconnected subtree $(\mathcal{V}_{\text{sub}}, \mathcal{E}_{\text{sub}})$ is reintroduced into the network by connecting its root node $n_{\text{sub}} \in \mathcal{V}_{\text{sub}}$ to an existing node $p \in \mathcal{V}_{\text{base}}$, ensuring that no cycles are created:
$\mathcal{V}_q = \mathcal{V}_{\text{base}} \cup \mathcal{V}_{\text{sub}}$, $\mathcal{E}_q = \mathcal{E}_{\text{base}} \cup \mathcal{E}_{\text{sub}} \cup \{ (n_{\text{sub}}, p) \}$.
\end{itemize}


In real-world distribution systems, reconfigurations frequently involve simultaneous operations such as multiple feeder disconnections and new feeder connections due to load balancing, fault isolation, network expansion, or emergency restoration, as illustrated in Fig.~\ref{transfer_gcn2}. These complex scenarios underscore the importance of considering multiple reconfiguration types at once for comprehensive augmentation strategies.

\begin{remark}
\textbf{Transmission vs. Distribution System Reconfigurations}: Transmission and distribution systems exhibit fundamentally different reconfiguration dynamics. Transmission systems experience infrequent topology changes primarily during maintenance or planned operations, while their meshed structure enables flexible power rerouting and provides inherent redundancy. 
Given the greater topological stability of transmission networks, we treat them as a specialized case within our augmentation framework, focusing primarily on line outages and scheduled maintenance scenarios. The more challenging transferability problem involves training a single UGCN across multiple disparate transmission networks (e.g., IEEE 30, 57, and 57 bus systems) to achieve robust generalization to entirely unseen operational reconfigurations.
\end{remark}

Next, we introduce grid-graph sampling as a novel approach for training graph convolutions across heterogeneous power networks.

 
\subsection{Grid-Graph Sampling}
 Grid-Graph Sampling is foundation of cross-system transfer learning in our work, enabling GCNs to learn from diverse grid topologies and operational data. 

More specifically, consider $Q$ distinct distribution systems, where each system $q \in \mathcal{Q} = \{1, 2, \ldots, Q\}$ represents an independent power network with unique topology and operational characteristics. For each power system $q$, the graph $\mathcal{G}_q = (\mathcal{V}_q, \mathcal{E}_q)$ consists of a node set $\mathcal{V}_q$ with $N_q = |\mathcal{V}_q|$ nodes, a weighted graph shift operator $\mathbf{S}_q \in \mathbb{C}^{N_q \times N_q}$, and a node feature matrix $\mathbf{X}_q \in \mathbb{C}^{N_q \times F_0}$, where $F_0$ represents the number of input node features.
The grid-graph sampling strategy involves randomly selecting different power systems $q \sim \mathcal{Q}$ during each training iteration, exposing the UGCN to diverse network topologies, varying system scales, and different operational patterns. 

For each power system $q \in \mathcal{Q}$, the graph convolution operation with multi-channel features is expressed as:
\begin{equation}
\mathbf{X}_q^{(\ell+1)} = \sigma\left(\sum_{k=0}^{K} \sum_{\tau=0}^{K_t} \mathbf{S}_q^k \mathbf{X}^{(\ell)}_{q, t-\tau} \mathbf{H}_{k,\tau}^{(\ell)}\right)
\end{equation}
where $\mathbf{X}_q^{(\ell)} \in \mathbb{C}^{N_q \times F_\ell}$ represents the node features for system $q$ at layer $\ell$, $\sigma(\cdot)$ is the activation function, and $\mathbf{H}_{k,\tau}^{(\ell)} \in \mathbb{C}^{F_\ell \times F_{\ell+1}}$ are the learnable weight matrices that remain constant across all systems, consistent with the spatio-temporal filter coefficients from the preliminaries.

The key insight is that the filter coefficients $\mathbf{H}_{k,\tau}^{(\ell)}$ are shared across all power systems, enabling direct transferability. While each system has its unique graph shift operator $\mathbf{S}_q$ and node features $\mathbf{X}_q$, the learned convolution parameters capture universal power system relationships.
The training objective function across all $Q$ power systems is:
\begin{equation}\label{eq:grid_sampling_objective}
\min_{\mathbf{H}} \mathcal{L}(\mathbf{H}) := \frac{1}{Q} \sum_{q=1}^Q \ell(\mathbf{H}; \mathbf{S}_q, \mathbf{X}_q, \mathbf{y}_q)
\end{equation}
where $\mathbf{y}_q$ represents the ground truth labels for system $q$, $\ell(\cdot)$ is the loss function for individual systems, and $\mathbf{H} = \{\mathbf{H}_{k,\tau}^{(\ell)}\}_{\ell=1}^L$ denotes the collection of all learnable parameters across layers.

During each training iteration, we sample a batch of power systems $\mathcal{B} \subset \mathcal{Q}$ and update the shared parameters $\mathbf{H}$ based on the aggregated loss across the sampled systems. This approach ensures that the UGCN  generalize across diverse power network configurations.
 
 \begin{figure*}[!htp]
 \vspace{-0.2cm}
  \center
    \includegraphics[width=0.75\textwidth]{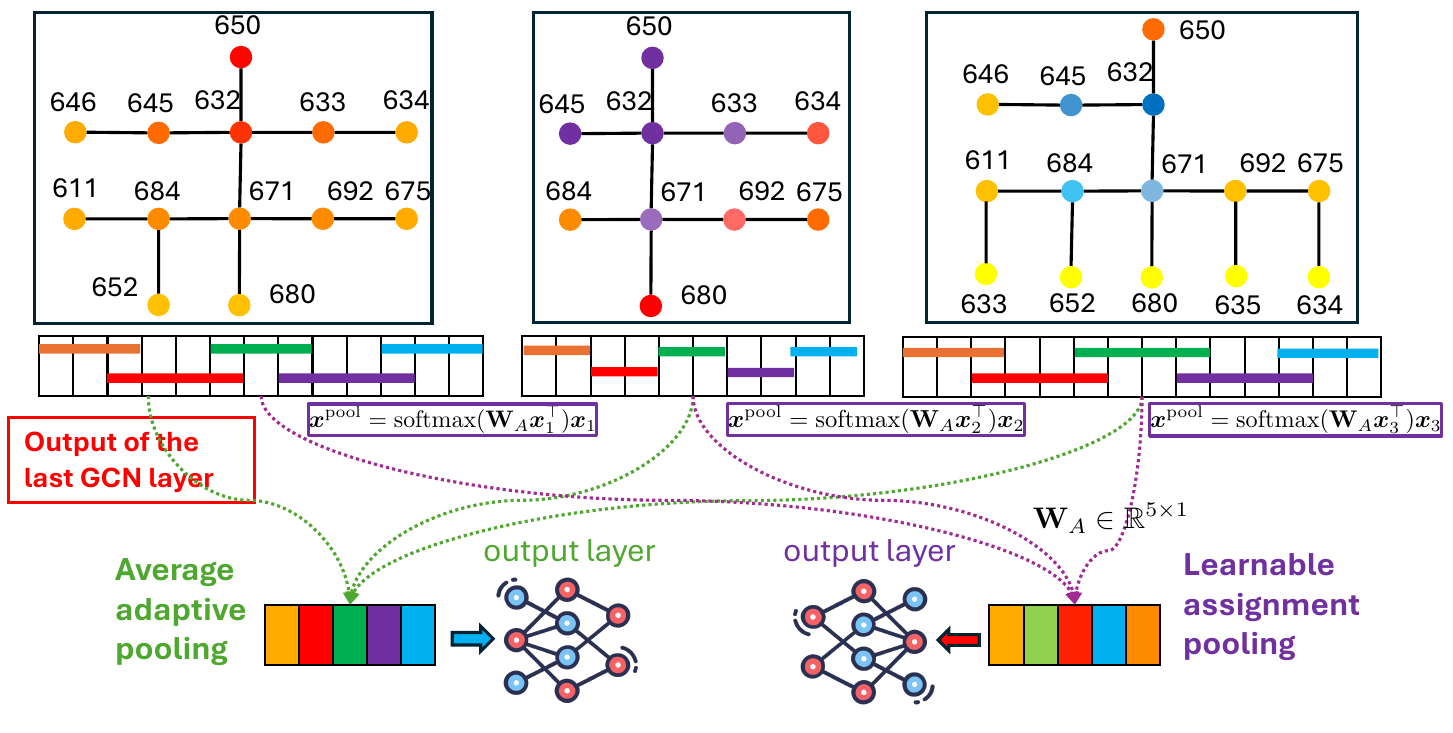}
      \vspace{-0.4cm}
  \caption{The output of the last GCN layer is processed using either average adaptive grid pooling or learnable assignment grid pooling. The former is faster and computes the average of the graph signal, while the latter is a trainable process. After adaptive grid pooling, the output attains a fixed dimension regardless of the varying input graph sizes, ensuring a consistent structure for the output layer despite differences in input graphs.}\label{adaptive_pool}
  \vspace{-0.6cm}
\end{figure*}

\subsection{Application-Oriented Adaptive Grid Pooling}
The grid-graph sampling strategy exposes the UGCN training to diverse network topologies, yet reconfiguration creates partially overlapping or entirely distinct feature distributions with varying dimensions. These distributions share common patterns when viewed through application-specific lenses. This subsection introduces application-oriented adaptive pooling to handle both input feature distribution shifts and dimensional mismatches through strategic information extraction rather than forced alignment.

In grid-graph sampling, different power systems $q \in \mathcal{Q}$ have varying numbers of nodes $N_q$, resulting in node feature matrices $\mathbf{X}_q^{(\ell+1)} \in \mathbb{C}^{N_q \times F_{\ell+1}}$ of different dimensions. This variability creates incompatibility issues when feeding the GCN outputs to subsequent fully connected layers, which require fixed input dimensions.

To address this challenge, we introduce \emph{Adaptive Grid Pooling}, a mechanism that dynamically transforms variable-size node representations $\mathbf{X}_q^{(\ell+1)}$ into fixed-size representations while preserving essential topological and feature information across different power systems.
We define an adaptive pooling matrix $\mathbf{A}_q \in \mathbb{C}^{N_p \times N_q}$ for each system $q$, where:
\begin{itemize}
    \item $N_q = |\mathcal{V}_q|$ is the number of nodes in power system $q$, which varies across different networks
    \item $N_p$ is a predefined number of pooled nodes, ensuring consistent dimensions for subsequent layers
\end{itemize}

The adaptive pooling operation is expressed as:
\begin{equation}
    \mathbf{X}_q^{\text{pool}} = \mathbf{A}_q \mathbf{X}_q^{(\ell+1)}
\end{equation}
where $\mathbf{X}_q^{\text{pool}} \in \mathbb{C}^{N_p \times F_{\ell+1}}$ serves as the fixed-size representation that can be uniformly processed by subsequent layers, regardless of the original system size $N_q$. Here, $F_{\ell+1}$ is the feature dimension of the final GCN layer output.

\subsubsection{Designing the Adaptive Pooling Matrix $\mathbf{A}_q$}
To construct $\mathbf{A}_q$, we explore two methods: \emph{adaptive average pooling} and \emph{learnable assignment pooling}.

\paragraph{Adaptive Custom Pooling}
To enrich the representation while maintaining a fixed output size, we partition the $N_q$ nodes of $\mathcal{V}_q$ into $N_p$ disjoint clusters $\{\mathcal{C}_1, \dots, \mathcal{C}_{N_p}\}$ such that $\bigcup_{i=1}^{N_p} \mathcal{C}_i = \mathcal{V}_q$ and $\mathcal{C}_i \cap \mathcal{C}_j = \emptyset$ for $i \neq j$, using a deterministic assignment strategy (e.g., based on electrical distance or topological ordering). For each cluster $\mathcal{C}_i$, we compute both the average and max pooled features:
\begin{align}
    \bm{z}^{\text{avg}}_i &= \frac{1}{|\mathcal{C}_i|} \sum_{n \in \mathcal{C}_i} \mathbf{X}_q^{(\ell+1)}[n,:], \\
    \bm{z}^{\text{max}}_i &= \max_{n \in \mathcal{C}_i} \mathbf{X}_q^{(\ell+1)}[n,:].
\end{align}
where $\mathbf{X}_q^{(\ell+1)}[n,:] \in \mathbb{C}^{F_{\ell+1}}$ denotes the feature vector of node $n \in \mathcal{V}_q$. Since each node's feature vector is in $\mathbb{C}^{F_{\ell+1}}$, both $\bm{z}^{\text{avg}}_i \in \mathbb{C}^{F_{\ell+1}}$ and $\bm{z}^{\text{max}}_i \in \mathbb{C}^{F_{\ell+1}}$. These two representations are concatenated to form the final feature vector for cluster $i$:
\begin{equation}
    \bm{z}_i = \left[ \bm{z}^{\text{avg}}_i \;\middle\| \; \bm{z}^{\text{max}}_i \right] \in \mathbb{C}^{2F_{\ell+1}}.
\end{equation}
The final pooled representation stacks all cluster features: $\mathbf{X}_q^{\text{pool}} = [\bm{z}_1; \bm{z}_2; \ldots; \bm{z}_{N_p}] \in \mathbb{C}^{N_p \times 2F_{\ell+1}}$.
In matrix form, average pooling can be written compactly as:
\begin{equation}
    \mathbf{Z}^{\text{avg}} = \mathbf{A}_q^{\text{(avg)}} \mathbf{X}_q^{(\ell+1)}, \quad
    [\mathbf{A}_q^{\text{(avg)}}]_{in} =
    \begin{cases}
        \frac{1}{|\mathcal{C}_i|}, & \text{if } n \in \mathcal{C}_i, \\
        0, & \text{otherwise}.
    \end{cases}
\end{equation}
where $\mathbf{A}_q^{\text{(avg)}} \in \mathbb{C}^{N_p \times N_q}$ and $\mathbf{Z}^{\text{avg}} \in \mathbb{C}^{N_p \times F_{\ell+1}}$. Similarly, max pooling produces $\mathbf{Z}^{\text{max}} \in \mathbb{C}^{N_p \times F_{\ell+1}}$, and the final pooled representation is $\mathbf{X}_q^{\text{pool}} = [\mathbf{Z}^{\text{avg}} \| \mathbf{Z}^{\text{max}}] \in \mathbb{C}^{N_p \times 2F_{\ell+1}}$.

This approach reduces the node dimension from $N_q$ to $N_p$ while doubling the feature dimension from $F_{\ell+1}$ to $2F_{\ell+1}$ through concatenation of average and max pooled features. For instance, with $N_p = 4$ pooled nodes, a 13-node system might use cluster sizes of $\{3, 4, 3, 3\}$, while a 10-node system uses $\{3, 3, 2, 2\}$, both producing 4 pooled nodes with $2F_{\ell+1}$ features each.

\paragraph{Learnable Assignment Pooling}
As an alternative to fixed averaging, we define a trainable assignment matrix $\mathbf{A}_q$, which allows the network to dynamically learn how nodes should be aggregated. This is formulated as:
\begin{equation}
    \mathbf{A}_q = \text{softmax}_{\text{rows}}(\mathbf{W}_A (\mathbf{X}_q^{(\ell+1)})^\top),
\end{equation}
where:
\begin{itemize}
    \item $\mathbf{W}_A \in \mathbb{C}^{N_p \times F_{\ell+1}}$ is a learnable weight matrix that projects node features into assignment scores
    \item $\text{softmax}_{\text{rows}}$ ensures each row of $\mathbf{A}_q$ sums to 1, meaning each pooled node receives contributions from multiple original nodes based on learned importance weights
    \item The transpose $(\mathbf{X}_q^{(\ell+1)})^\top \in \mathbb{C}^{F_{\ell+1} \times N_q}$ ensures proper matrix multiplication compatibility
\end{itemize}

The resulting assignment matrix $\mathbf{A}_q \in \mathbb{R}^{N_p \times N_q}$ adapts its dimensions based on the input network size. For example, if the input is a feature matrix $\mathbf{X}_q^{(\ell+1)} \in \mathbb{C}^{13 \times F_{\ell+1}}$ from a 13-node system or $\mathbf{X}_q^{(\ell+1)} \in \mathbb{C}^{10 \times F_{\ell+1}}$ from a 10-node system, the weight matrix $\mathbf{W}_A$ remains fixed at $N_p \times F_{\ell+1}$. Consequently, $\mathbf{A}_q$ is adaptively sized as $N_p \times 13$ or $N_p \times 10$, ensuring compatibility across different system configurations.

\subsubsection{Final Integration with GCN and FNN}
Applying $\mathbf{A}_q$ ensures that the hidden representations from the GCN are mapped to a fixed-size representation:
\begin{equation}
    \mathbf{X}_q^{\text{pool}} = \mathbf{A}_q \mathbf{X}_q^{(\ell+1)} \in \mathbb{C}^{N_p \times F_{\ell+1}}
\end{equation}
This facilitates seamless integration with the FNN output layer, which produces the final task-specific outputs $\bm{y}_q$. The choice between adaptive average pooling and learnable assignment pooling balances efficiency and flexibility. Adaptive average pooling offers a straightforward, deterministic approach, while learnable assignment pooling enables adaptive feature aggregation, enhancing performance in complex graph structures.

Having dealt withinput variability, in the next section we describe the UGCN framework to manage output variations. 

\subsection{Adaptive Parallel Transformer Output}
Adaptive pooling resolves dimensional variability during feature extraction, but output generation poses a different challenge: producing variable-sized outputs while remaining efficient. To address this, we introduce a parallel Transformer architecture that generates outputs of varying dimensions through parallel processing.

For a power network with $N_q$ buses, the Transformer receives the pooled features $\mathbf{X}_q^{pool} \in \mathbb{R}^{N_p \times F_L}$ from the adaptive pooling layer. These features are first flattened and encoded into a unified representation:
\begin{equation}
\mathbf{h}_q = \sigma(\mathbf{W}_{enc} \text{vec}(\mathbf{X}_q^{pool}) + \mathbf{b}_{enc}) \in \mathbb{R}^{d}
\end{equation}
where $\text{vec}(\mathbf{X}_q^{pool}) \in \mathbb{R}^{N_p \cdot F_L}$ flattens the pooled features, $d$ is the hidden dimension, $\mathbf{W}_{enc} \in \mathbb{R}^{d \times (N_p \cdot F_L)}$ is the encoding weight matrix, $\mathbf{b}_{enc} \in \mathbb{R}^{d}$ is the encoding bias, and $\sigma(\cdot)$ is the activation function (typically ReLU).
The key innovation lies in parallel position encoding that generates embeddings for all $N_q$ output positions simultaneously:
\begin{align}
\mathbf{p}_{N_q} &= \left[\frac{0}{\max(N_q-1,1)},  \ldots, \frac{N_q-1}{\max(N_q-1,1)}\right]^T \in \mathbb{R}^{N_q \times 1} \\
\mathbf{E}_{pos,q} &= \tanh(\mathbf{W}_{pos} \mathbf{p}_{N_q} + \mathbf{B}_{pos}) \in \mathbb{R}^{N_q \times d}
\end{align}
where $\mathbf{W}_{pos} \in \mathbb{R}^{d \times 1}$ is the position encoding weight matrix and $\mathbf{B}_{pos} \in \mathbb{R}^{N_q \times d}$ represents the position bias.

The encoded features are then broadcast and combined with position embeddings through parallel transformation:
\begin{align}
\mathbf{C}_q &= \mathbf{1}_{N_q} \mathbf{h}_q^T + \mathbf{E}_{pos,q} \in \mathbb{R}^{N_q \times d} \\
\mathbf{T}_q &= \sigma(\mathbf{C}_q \mathbf{W}_T + \mathbf{1}_{N_q} \mathbf{b}_T^T) \in \mathbb{R}^{N_q \times d} \\
\mathbf{y}_q &= \sigma(\mathbf{T}_q \mathbf{w}_{out} + b_{out}) \in \mathbb{R}^{N_q}
\end{align}
where $\mathbf{1}_{N_q} \in \mathbb{R}^{N_q}$ is a vector of ones, $\mathbf{W}_T \in \mathbb{R}^{d \times d}$ is the transformation weight matrix, $\mathbf{b}_T \in \mathbb{R}^{d}$ is the transformation bias, $\mathbf{w}_{out} \in \mathbb{R}^{d}$ is the output weight vector, and $b_{out} \in \mathbb{R}$ is the output scalar bias.

This parallel adaptive mechanism provides three critical advantages: \textbf{(1) Universal weight reuse} - the same parameter set $\{\mathbf{W}_{enc}, \mathbf{W}_{pos}, \mathbf{W}_T, \mathbf{w}_{out}, \mathbf{b}_{enc}, \mathbf{B}_{pos}, \mathbf{b}_T, b_{out}\}$ handles networks of any size   through learned position-to-embedding mappings; \textbf{(2) Computational efficiency} - parallel processing achieves $O(N_q \cdot d^2)$ complexity compared to $O(N_q^2 \cdot d^2)$ for autoregressive approaches; \textbf{(3) Order-agnostic adaptivity} - the model processes networks in any sequence (e.g., training $57 \to 30 \to 39$, testing $30 \to 57 \to 39$) since each network size is handled independently through dynamic position matrix generation.  

\begin{remark}
    Our adaptive output mechanism preserves the loss function structure and task semantics while adapting to topology-induced variations in how and where network elements contribute to the loss. As networks reconfigure, the computational paths, contributing buses, and parameter distributions change, but the fundamental optimization objective remains consistent.
\end{remark}

\begin{figure}[!htbp]
\vspace{-0.2cm}
\centering
\includegraphics[width=0.4\textwidth]{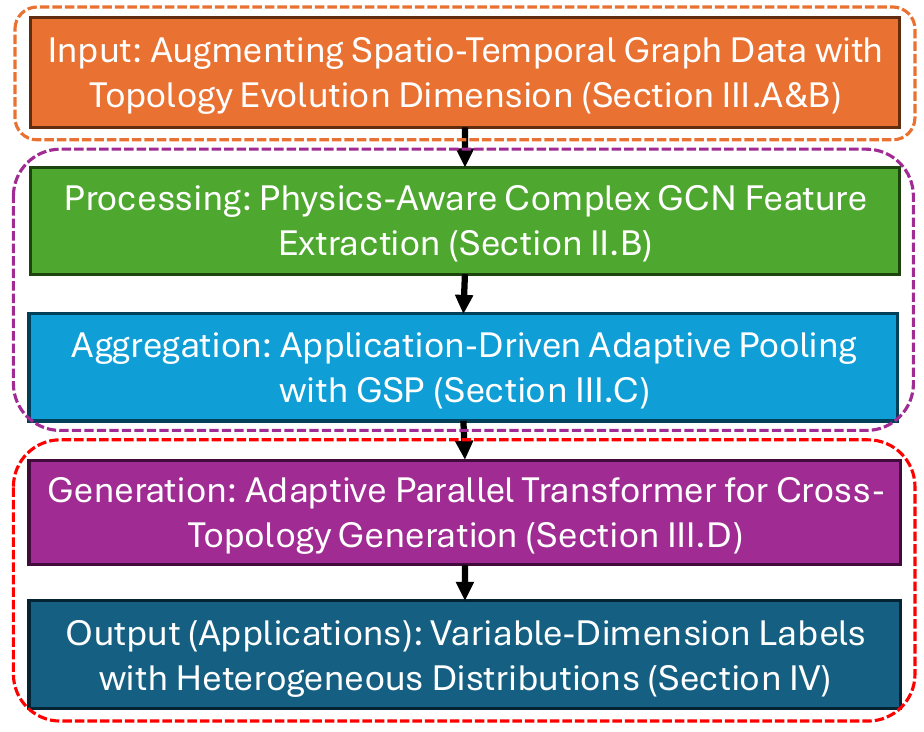}
\caption{UGCN framework overview: From problem formulation through universal graph learning architecture to cross-system applications.}
\label{fig:flowchart}
\vspace{-0.6cm}
\end{figure}

\subsection{Summary of UGCN Algorithm Design}
As shown in Algorithm~\ref{alg_UGCN}, the UGCN framework enables generalization across different grid topologies through graph augmentation, spatio-temporal convolution, adaptive pooling, and adaptive Transformer output. Figure~\ref{fig:flowchart} illustrates the complete methodology workflow. The process: (1) augment base graph into multiple system variations, (2) train with shared spatio-temporal parameters across systems, (3) use adaptive pooling or Transformer for variable-size outputs, (4) deploy on unseen networks via parameter transfer.

\begin{algorithm}[!htb] \small
\caption{UGCN Training and Inference Framework}\label{alg_UGCN}
\KwIn{Base graph $\mathcal{G}_{\text{base}} = (\mathcal{V}_{\text{base}}, \mathcal{E}_{\text{base}})$, 
Features $\mathbf{X}_{\text{base}} \in \mathbb{C}^{N_{\text{base}} \times F_0}$, GSO $\mathbf{S}_{\text{base}} \in \mathbb{C}^{N_{\text{base}} \times N_{\text{base}}}$,  
$Q$ systems, $T$ epochs, learning rate $\eta$, pooled nodes $N_p$, output method $\in \{\text{pooling}, \text{transformer}\}$}
\KwOut{Trained UGCN parameters $\mathbf{H} = \{\mathbf{H}_{k,\tau}^{(\ell)}\}_{\ell=1}^L$ and output weights}

\textbf{Step 1: Graph Augmentation} \\
\For{$q = 1$ \textbf{to} $Q$}{
    Generate $\mathcal{G}_q = (\mathcal{V}_q, \mathcal{E}_q)$ via feeder ops, line breaks, parameter changes \\
    Compute $\mathbf{S}_q \in \mathbb{C}^{N_q \times N_q}$, initialize $\mathbf{X}_q^{(0)} = \bm{v}_q \in \mathbb{C}^{N_q \times F_0}$ \\
}
\textbf{Step 2: Training with Shared Parameters} \\
Initialize $\mathbf{H}_{k,\tau}^{(\ell)} \in \mathbb{C}^{F_\ell \times F_{\ell+1}}$ and output weights randomly \\
\For{epoch = 1 \textbf{to} $T$}{
    Sample batch $\mathcal{B} \subset \mathcal{Q}$ \\
    \For{\textbf{each} $q \in \mathcal{B}$}{
        \For{$\ell = 1$ \textbf{to} $L$}{
            $\mathbf{X}_q^{(\ell)} = \sigma\left(\sum_{k=0}^{K} \sum_{\tau=0}^{K_t} \mathbf{S}_q^k \mathbf{X}^{(q, \ell-1)}_{t-\tau} \mathbf{H}_{k,\tau}^{(\ell)}\right)$ \\
        }
        Apply adaptive pooling: $\mathbf{X}_q^{\text{pool}} = \mathbf{A}_q \mathbf{X}_q^{(L)} \in \mathbb{C}^{N_p \times F_L}$ \\
        \eIf{transformer output}{
            $\bm{h}_q = \sigma(\mathbf{W}_{enc} \text{mean}(\mathbf{X}_q^{\text{pool}}) + \bm{b}_{enc}) \in \mathbb{C}^{d_h}$ \\
            $\bm{p}_{N_q} = [0/(N_q-1), 1/(N_q-1), \ldots, (N_q-1)/(N_q-1)]^T$ \\
            $\mathbf{E}_{pos,q} = \tanh(\mathbf{W}_{pos} \bm{p}_{N_q}) \in \mathbb{C}^{N_q \times d_h}$ \\
            $\mathbf{C}_q = \bm{h}_q \otimes \bm{1}_{N_q}^T + \mathbf{E}_{pos,q}$ \\
            $\bm{y}_q = \sigma(\mathbf{C}_q \mathbf{W}_T + \bm{b}_T) \mathbf{W}_{out} + \bm{b}_{out} \in \mathbb{C}^{N_q}$ \\
        }{
            $\bm{y}_q = \sigma(\mathbf{W} \mathbf{X}_q^{\text{pool}}) \in \mathbb{C}^{N_p}$ \\
        }
    }
    $\mathcal{L}(\mathbf{H}) = \frac{1}{|\mathcal{B}|} \sum_{q \in \mathcal{B}} \ell(\mathbf{H}; \mathbf{S}_q, \mathbf{X}_q, \mathbf{y}_q)$ \\
    Update all parameters: $\mathbf{H}, \mathbf{W}_{enc}, \mathbf{W}_{pos}, \mathbf{W}_T, \mathbf{W}_{out}, \ldots$ \\
}
\textbf{Step 3: Inference on New Systems} \\
\For{\textbf{each} $\mathcal{G}_{\text{new}} = (\mathcal{V}_{\text{new}}, \mathcal{E}_{\text{new}})$}{
    Compute $\mathbf{S}_{\text{new}} \in \mathbb{C}^{N_{\text{new}} \times N_{\text{new}}}$, $\mathbf{X}_{\text{new}}^{(0)} = \bm{v}_{\text{new}}$ \\
    Apply trained UGCN: $\bm{y}_{\text{new}} = \text{UGCN}(\mathbf{S}_{\text{new}}, \mathbf{X}_{\text{new}}^{(0)}; \mathbf{H})$ \\
}
\end{algorithm}


\section{UGCN Applications in Power Systems}
This section explores two key applications of the UGCN in power systems: forecasting power system state variables and detecting false data injection (FDI) attacks.   

\subsection{Power Distribution System State Forecasting}
In the GSP-based Nonlinear System State Estimation (NSSE) formulation, the objective is to estimate the complex voltage phasors $\bm{v} \in \mathbb{C}^{N}$ at each bus while incorporating the nonlinear nature of AC power flow equations. The measurement model is given by  
\begin{equation}
\bm{z} = h(\bm{v}) + \bm{\varepsilon}, \label{eq:measurement_model}
\end{equation}
where $h(\bm{v})$ represents the nonlinear mapping from bus voltages to measured quantities such as power injections, power flows, and voltage magnitudes, and $\bm{\varepsilon}$ is measurement noise. The power injection at bus $n \in \mathcal{V}$ is expressed as  
\begin{equation}
S_n = P_n + \mathfrak{j} Q_n = v_n \sum_{m \in \mathcal{N}(n)} Y_{nm}^* v_m^*, \label{eq:power_injection}
\end{equation}
where $Y_{nm}$ is the complex admittance matrix element, and $v_m^*$ denotes the complex conjugate of $v_m$. 

\subsubsection{Scenario 1: Fast State Estimation}
This problem objective is a weighted least squares regularized cost:
\begin{equation}
\min_{\bm{v}_t} \left\| \bm{R}^{1/2} (\bm{z} - h(\bm{v}_t)) \right\|^2 + \lambda \left\| \bm{\Lambda}^{1/2} \bm{v}_t \right\|^2, \label{eq:distflow_sse}
\end{equation}
where $\bm{v}_t = [v_{1,t}, \ldots, v_{N,t}]^T \in \mathbb{C}^N$ represents the state vector containing complex voltage phasors at all buses at time $t$. The measurement function $h(\bm{v}_t)$ relates the voltage states to the available measurements $\bm{z}$ through nonlinear power flow equations. The weight matrix $\bm{R}$ accounts for measurement uncertainties, while $\bm{\Lambda}$ provides regularization.

The optimization can be solved using various approaches such as Newton's method or linearized techniques, yielding the state estimate that best fits the available measurements while satisfying the network model constraints. In this implementation,  variations of DistFlow equations are employed to enable fast computations for distribution network state estimation.

\begin{figure*}[!htbp]
\centering
\begin{minipage}{1.00\textwidth}
    \centering
    \subfigure[Traditional NN: Temporal training on IEEE 33-bus system with 2D Isomap visualization of high-dimensional input features]{\includegraphics[width=0.33\textwidth]{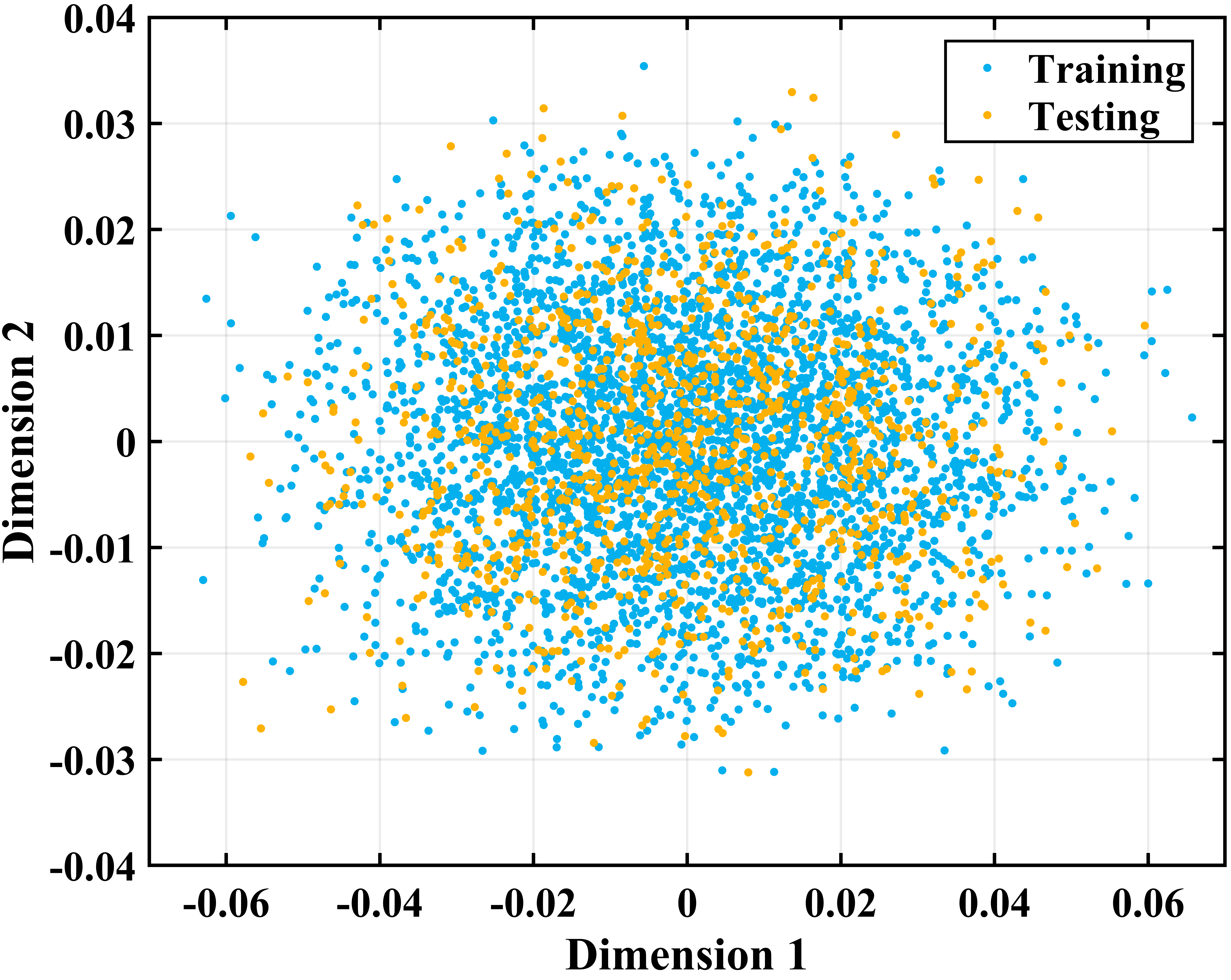}\label{fig:traditional_nn}}
    \hspace{-0.05in}
    \vspace{-0.05in}   
    \subfigure[UGCN: Cross-reconfiguration learning on IEEE 33-bus with 3D Isomap visualization across 1200 network reconfigurations]{\includegraphics[width=0.34\textwidth]{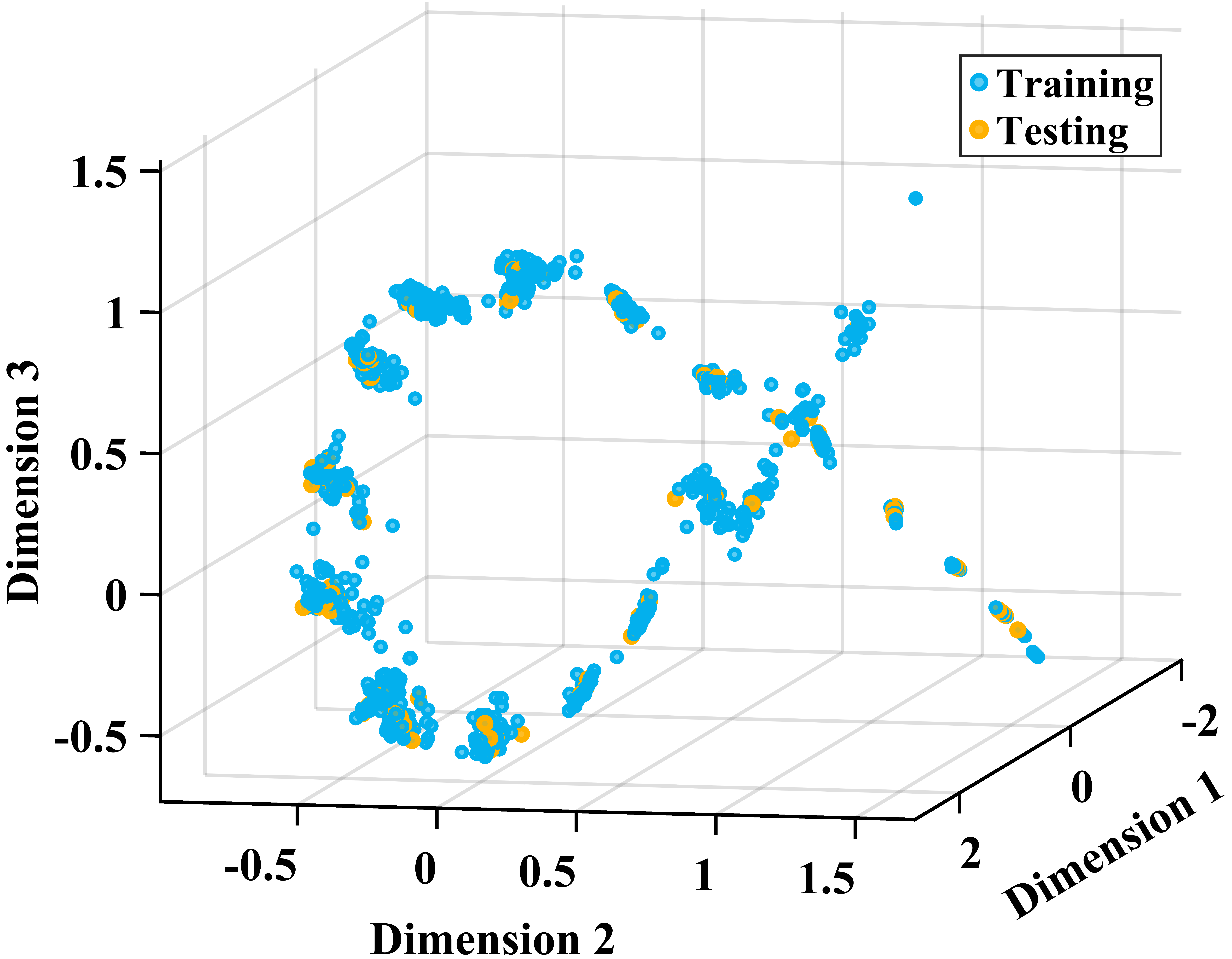}\label{fig:ugcn_reconfig}}
    \hspace{-0.05in}
    \vspace{-0.05in}     
    \subfigure[Universal challenge: IEEE 39-bus and IEEE 57-bus   systems exhibit distinct manifold structures even in low-dimensional projections]{\includegraphics[width=0.30\textwidth]{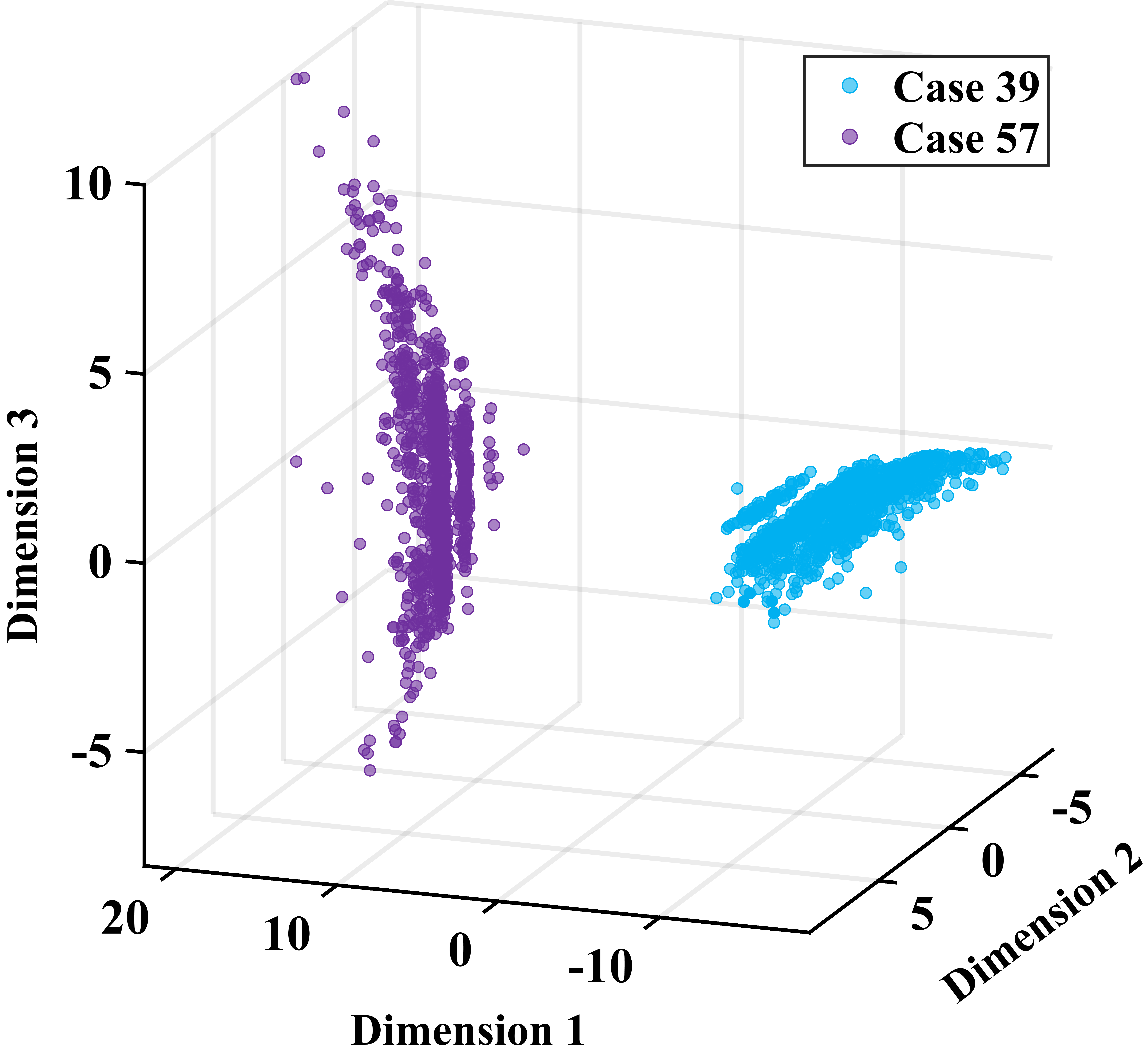}\label{fig:cross_system}}
    \caption{Visualization of data manifold complexity demonstrating the universal learning challenge: (a) Traditional temporal learning within a single configuration, (b) UGCN's ability to handle reconfigurations within one system topology, and (c) the fundamental challenge requiring a single UGCN to simultaneously cover both IEEE 39-bus and IEEE 57-bus transmission systems with their respective reconfigurations.}
    \label{fig:manifold_analysis}
\end{minipage}
\vspace{-0.6cm}
\end{figure*}

\subsubsection{Scenario 2: Sparse Deployment PMU-Based NSSE}
Given the selected PMU locations, the measurement model \eqref{eq:measurement_model} is reformulated as:
\begin{equation}
\bm{z}_t =
\bm{H} \bm{v}_t + \bm{\varepsilon}_t, \quad
\bm{H} = \begin{bmatrix}
\mathbf{Y}_{\mathcal{A} \mathcal{A}} & \mathbf{Y}_{\mathcal{A} \mathcal{U}} \\
\mathbb{I}_{|\mathcal{A}|} & \mathbf{0}
\end{bmatrix}, \label{eq:pmu_measurement}
\end{equation}
where $\bm{x}_t \in \mathbb{C}^N$ represents the system-wide voltage phasors at time $t$, $\mathcal{A}$ denotes the set of buses with PMUs, and $\mathcal{U}$ represents unobserved buses. The estimated state $\hat{\bm{x}}_t$ is obtained using the regularized least squares approach:
\begin{equation}
  \hat{\bm{x}}_t = \left(\bm{H}^\mathsf{H} \bm{H}  + \mu_1 \mathbf{S} \right)^\dagger \bm{H}^\mathsf{H} \bm{z}_t. \label{eq:state_estimation}
\end{equation}
The estimated phasor at bus $n$ is then: $\hat{x}_{t, n} = [\hat{\bm{x}}_t]_n. $ 
 
\begin{tcolorbox}[breakable]
Given the estimated state at $t$, irrespective of the measurement system,  the goal of the UGCN is to learn how to perform a maximum a posteriori probability prediction of the phasor at bus $n$ at time $t+H$, i.e.:
\begin{equation}
\hat{x}_{t+H, n} = \underset{x_{t+H, n}}{\arg \max } P \left( x_{t+H, n} \mid \hat{\mathbf{X}}_{t, n} \right), \label{eq:forecasting}
\end{equation}
where the input feature vector is:
\begin{equation}
\hat{\mathbf{X}}_{t, n} = \left[ \hat{x}_{t-T+1:t, n}, \hat{x}_{t-T+1:t, \mathcal{N}(n)} \right]. \label{eq:forecasting_input}
\end{equation}
\end{tcolorbox}

\subsection{False Data Injection Detection}
The second application of UGCN is detecting FDI attacks, where the attacked measurements are modeled as:
\begin{equation}
\bm{z}_t = \bm{H} (\bm{v}_t + \omega \cdot \delta \bm{v}_t) + \bm{\varepsilon}_t. \label{eq:attack_model}
\end{equation}
where $\delta \bm{v}_t$ is the perturbation, $\omega$ is a scaling factor from $[0, 1]$ to adjust the attack magnitude. Smaller values of $\omega$ result in less system impact but make attacks harder to detect, creating a trade-off between attack stealth and effectiveness.
A type of attack known as a stealth attack involves an attacker manipulating current and voltage phasor measurements on a subset of buses, denoted as $\mathcal{C}$, by introducing a perturbation:
\begin{equation}
\begin{aligned}
\delta \bm{v}_{t}^{\top}=\left[\begin{array}{ll}
\delta \bm{v}_{\mathcal{C}}^{\top} & \mathbf{0}_{\mathcal{P} \cup \mathcal{U} }^{\top}
\end{array}\right], \text{ such that } \mathbf{Y}_{\mathcal{P}\mathcal{C}} \delta \bm{v}_{\mathcal{C}}=\mathbf{0},   \mathcal{C}  \subset \mathcal{A}
\end{aligned}
\end{equation}
where $\mathcal{P}$ is the set of honest nodes.
\begin{tcolorbox}[breakable]
The model classifies whether bus $n$ has been compromised:
\begin{equation}
\hat{y}_{n,t} = \underset{y_{n,t}}{\arg \max } P \left( y_{n,t} \mid \hat{\mathbf{X}}_{n,t} \right). \label{eq:FDI_detection}
\end{equation}
The input feature vector is identical to \eqref{eq:forecasting_input}, capturing both historical estimated phasors and neighboring bus states. The ground-truth label vector is defined as:
\begin{equation}
    \bm{y}_n = \text{logit}(\delta \bm{v}_n), \label{eq:logit}
\end{equation}
where $\bm{y}_n = 1$ if bus $n$ is compromised ($[\delta \bm{v}]_n \neq 0$), and $\bm{y}_n = 0$ otherwise.
\end{tcolorbox}

By using the subgraph-based formulation, the same feature extraction process in \eqref{eq:forecasting_input} is leveraged for both forecasting and FDI detection. This enables robust voltage phasor predictions and attack detection even under sparse PMU deployment.

\section{Case Studies}
\subsection{Simulation Settings}
For the simulation, we train a UGCN using data from the IEEE 33-bus and 69-bus radial distribution systems. To extend the study to transmission systems, we include the IEEE 30-bus, IEEE 39-bus, and IEEE 57-bus networks, representing different regional power systems.
The power demand data is sourced from real-world measurements in Texas\footnote{\url{https://www.ercot.com/gridinfo/load/load_hist}}, while photovoltaic (PV) power data for smart inverters is obtained from the National Renewable Energy Laboratory (NREL)\footnote{\url{https://www.nrel.gov/grid/solar-power-data.html}}.

Reconfigurations in radial distribution systems can significantly alter network structure. For instance, the IEEE 33-bus network may vary between 22 and 38 nodes due to feeder disconnections, new feeder installations, link failures, subtree mergers, and simultaneous parameter changes. Our testing framework accounts for multiple reconfiguration events occurring randomly and concurrently, rather than in isolation. Similarly, the IEEE 30-bus transmission system can experience dynamic topology changes, including multiple link failures and parameter variations---without complete subgraph disconnection.
The testing dataset includes new reconfigurations and future time-series data unseen during training, assessing the UGCN's transferability across significant topological changes. For the power system state forecasting (PSSF) task, we use 1000 distribution system reconfigurations for training and 200 for testing. For FDI detection, we validate the UGCN on transmission systems with 1000 graphs for training and 200 reconfigurations for testing.
For distribution networks, the UGCN input consists of Advanced Metering Infrastructure (AMI) measurements, as PMU data is not widely deployed in these systems. We assume that AMI devices are installed at all feeder endpoints. For transmission networks, we use sparse time-series voltage phasor data from approximately 30\% of the buses. The input format includes $m_0 = 10$ channels, representing a multi-channel structure.

The UGCN architecture for the PSSF task includes two graph convolution layers for feature extraction, followed by two fully connected layers with 256 neurons each. For the FDI detection task, which focuses solely on spatial information, we adopt a similar architecture: one UGCN layer for spatial feature learning, followed by two fully connected layers with 256 neurons each.

\subsection{Manifold Structure: The Universal Learning Challenge}
 The low-dimensional representation of data manifold structures in Figure \ref{fig:manifold_analysis} reveals why developing a single universal UGCN model presents significant challenges. They are simplified projections from high-dimensional complex-valued feature spaces to 2D or 3D representations, where the actual separations and sparsity in the original high-dimensional space are far more pronounced than what these reduced projections can reveal \cite{meilua2024manifold, krishnan2018challenges}.

\begin{itemize}
\item \textbf{Single Configuration Learning (Figure \ref{fig:traditional_nn}):} Traditional neural networks operate effectively within the relatively dense spatio-temporal data from a single configuration, where the IEEE 33-bus system training data from the past year (blue) and testing data (orange) show closely overlapping distributions when mapped from 33 dimensions to 2 dimensions.

\item \textbf{Cross-Reconfiguration Learning (Figure \ref{fig:ugcn_reconfig}):} The sparse 3D manifold across 1200 reconfigurations within one system demonstrates the dramatically increased complexity when accommodating network topology changes. Each point represents an entire class of spatio-temporal distributions similar to those shown in Figure \ref{fig:traditional_nn}, yet these configuration classes are distributed sparsely across the manifold rather than forming dense, overlapping clusters, creating significant challenges for model generalization across reconfigurations.

\item \textbf{Universal Cross-System Learning (Figure \ref{fig:cross_system}):} Achieving universality requires simultaneously accommodating fundamentally different topologies, where IEEE 39-bus and IEEE 57-bus transmission systems occupy completely separate manifolds even in low-dimensional projections, illustrating the severe domain gap between different system architectures. This separation demonstrates why a single UGCN must bridge vast gaps between entirely different transmission topologies while maintaining separate yet coordinated representations for each system without catastrophic forgetting.
\end{itemize}

\begin{figure}[!htbp]
 \vspace{-0.2in}  
    \centering
    \subfigure[Transfer performance: AMI-based voltage forecasting]{\includegraphics[width=0.24\textwidth]{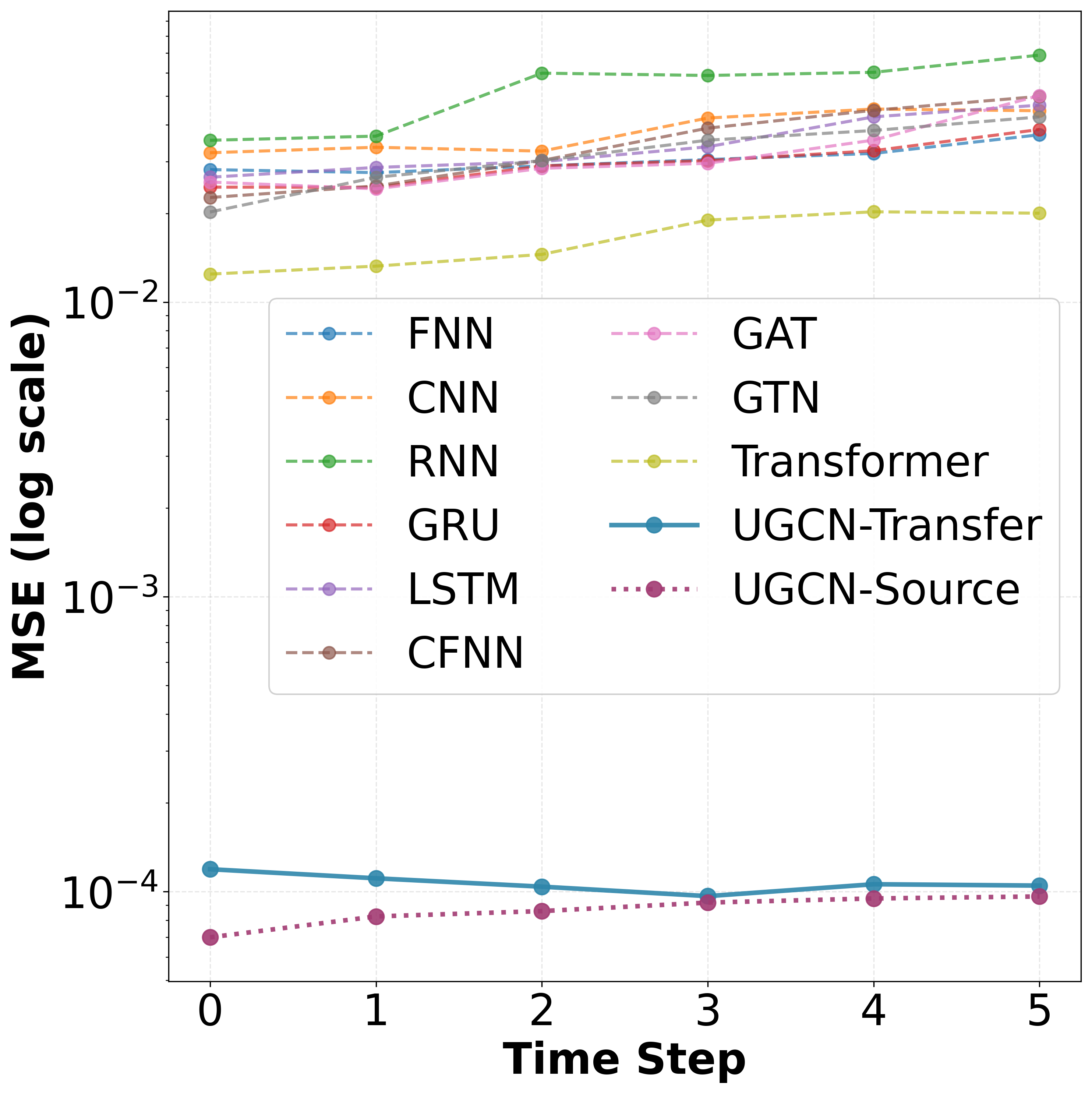}\label{fig:ami_33}}
    \hspace{-0.05in} 
       \vspace{-0.1in}  
    \subfigure[Transfer performance: PMU-based voltage forecasting]{\includegraphics[width=0.24\textwidth]{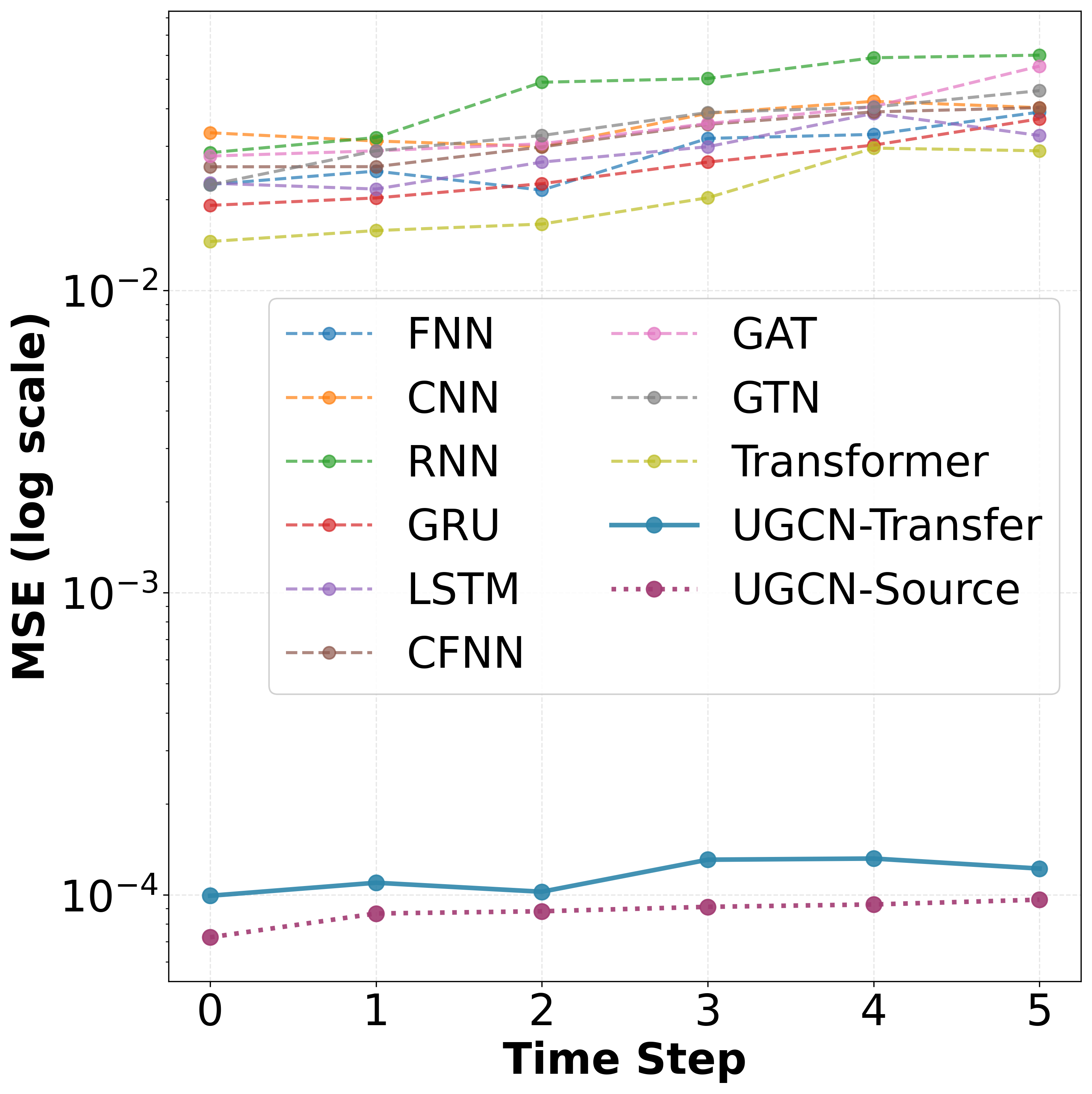}\label{fig:pmu_33}}
    \caption{UGCN transferability for power system state estimation and forecasting across unseen IEEE 33-bus reconfigurations}
    \label{fig:pssf33}
 \vspace{-0.4cm} 
\end{figure}

\begin{figure}[!htbp] 
    \centering
    \subfigure[Transfer performance: AMI-based voltage forecasting]{\includegraphics[width=0.24\textwidth]{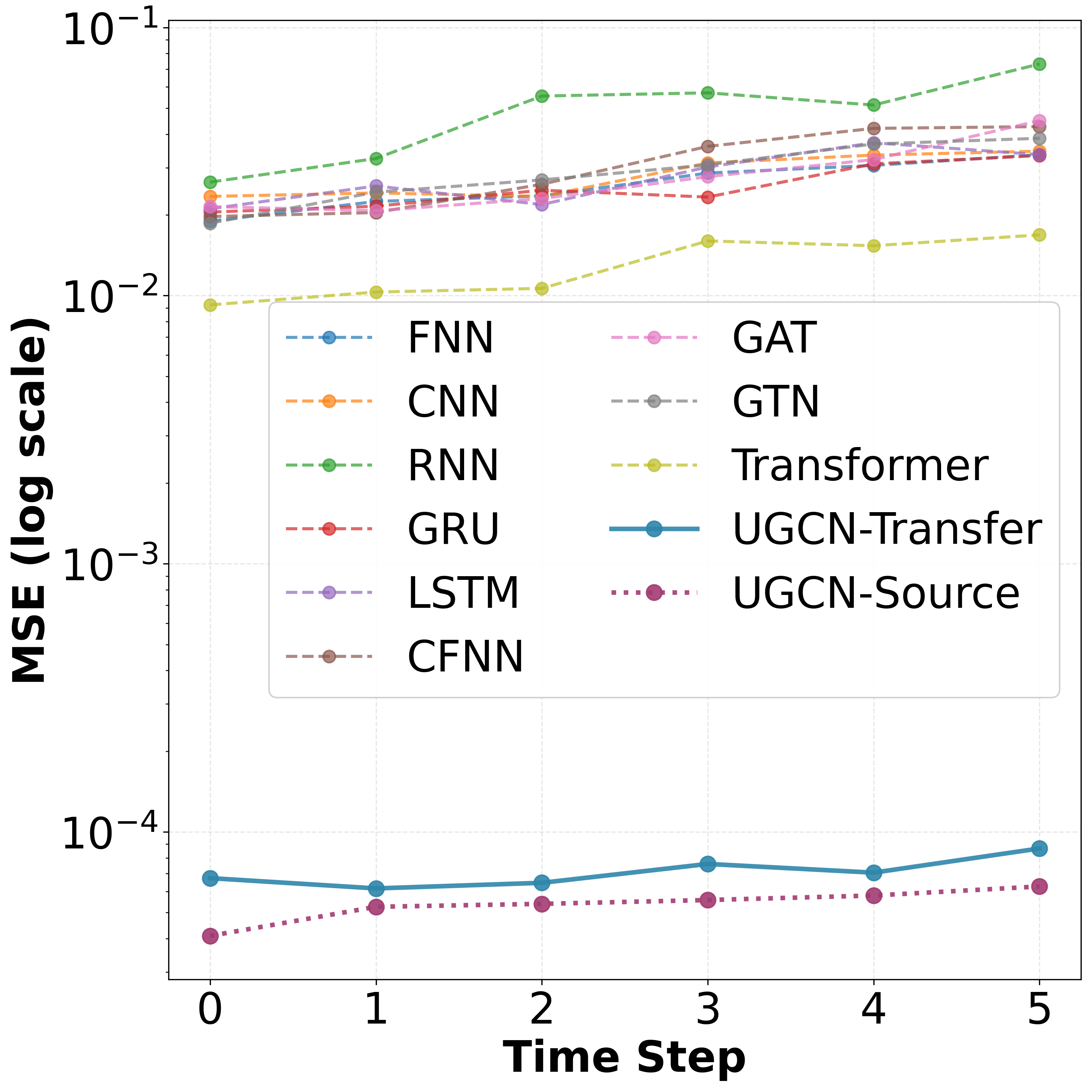}\label{fig:ami_69}}
    \hspace{-0.05in} 
       \vspace{-0.1in}  
    \subfigure[Transfer performance: PMU-based voltage forecasting]{\includegraphics[width=0.24\textwidth]{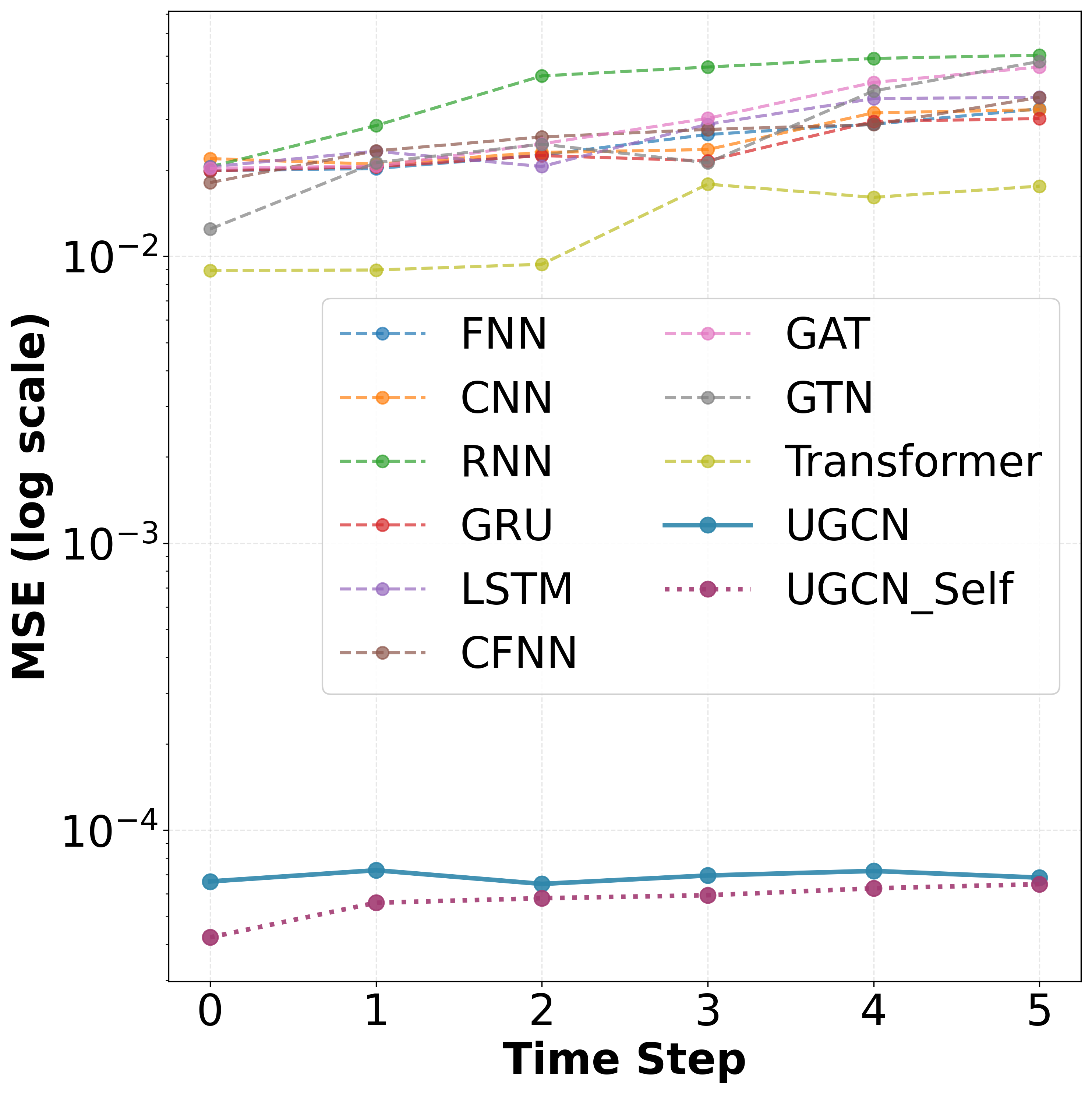}\label{fig:pmu_69}}
    \caption{UGCN transferability for power system state estimation and forecasting across unseen IEEE 69-bus reconfigurations}
    \label{fig:pssf69}
 \vspace{-0.4cm} 
\end{figure}

\subsection{Baseline Setting}
For both application domains, our baseline algorithms include \textbf{FNN} (four-layer fully-connected network with 512 neurons per layer using real/imaginary voltage parts), \textbf{CplxFNN} (complex-valued fully-connected network with identical architecture using complex voltage phasors), \textbf{GRU} (incorporating first-layer GNN \cite{kipf2017semi} followed by three fully-connected layers utilizing voltage amplitudes and angles), \textbf{Transformer} \cite{vaswani2017attention} (three sub-encoder layers in encoder/decoder sections with 512 features accommodating real/imaginary voltage phasor measurements), \textbf{GAT} \cite{velickovic2017graph} (graph attention layers with 10 input/output channels followed by three 512-neuron fully-connected layers processing voltage amplitudes and angles), \textbf{GTN} \cite{rampasek2022GPS} (similar architecture to GAT with graph transformer layers), \textbf{RNN} \cite{zhang2019power} (integrating voltage magnitudes, active/reactive power injections, and power flows\footnote{For each bus $n \in \mathcal{N}$, measurements comprise the voltage phasor $V_n:=|V_n| e^{\mathfrak{j} \theta_n}$ and complex power injection $S_n:=P_n+\mathfrak{j} Q_n$, complex power flows $S_{ij}^f:=P_{ij}^f+\mathfrak{j} Q_{ij}^f$ at the 'forwarding' end and $S_{ij}^e:=P_{ij}^e+\mathfrak{j} Q_{ij}^e$ at the 'terminal' end.}), \textbf{CNN} \cite{wang2020locational} (employing line and bus measurements with comparable sensor count), and \textbf{LSTM} \cite{wang2021kfrnn} (processing real/imaginary or complex-valued voltage phasors).

\begin{figure}[!htbp]
  \vspace{-0.3cm}
  \centering
  \subfigure[State estimation: Voltage magnitudes ($H=0$)]{
 \includegraphics[width=1.6in]{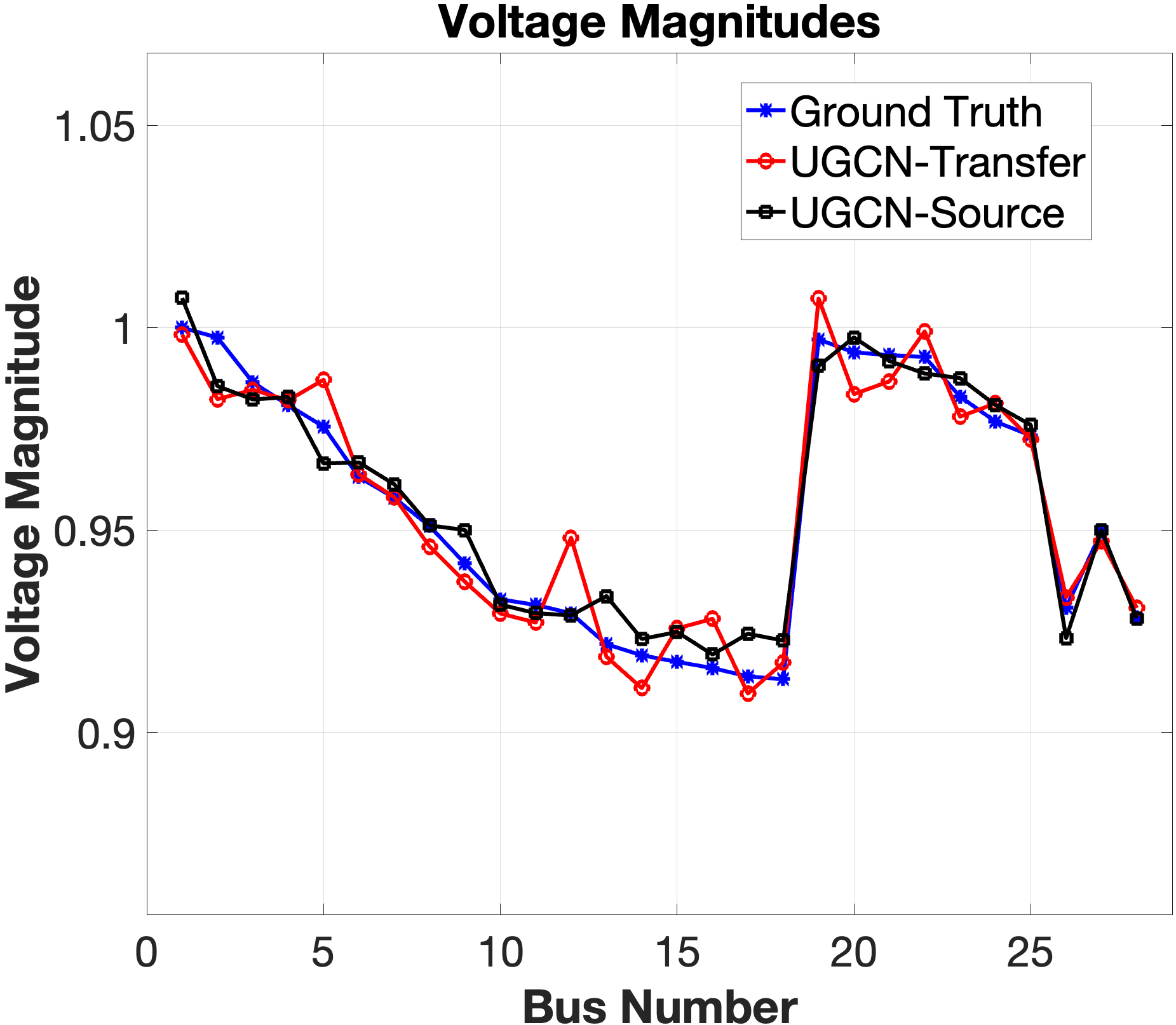}
      \label{fig:state_est_mag}
 } 
 \subfigure[State estimation: Voltage phase angles ($H=0$)]{
\includegraphics[width=1.6in]{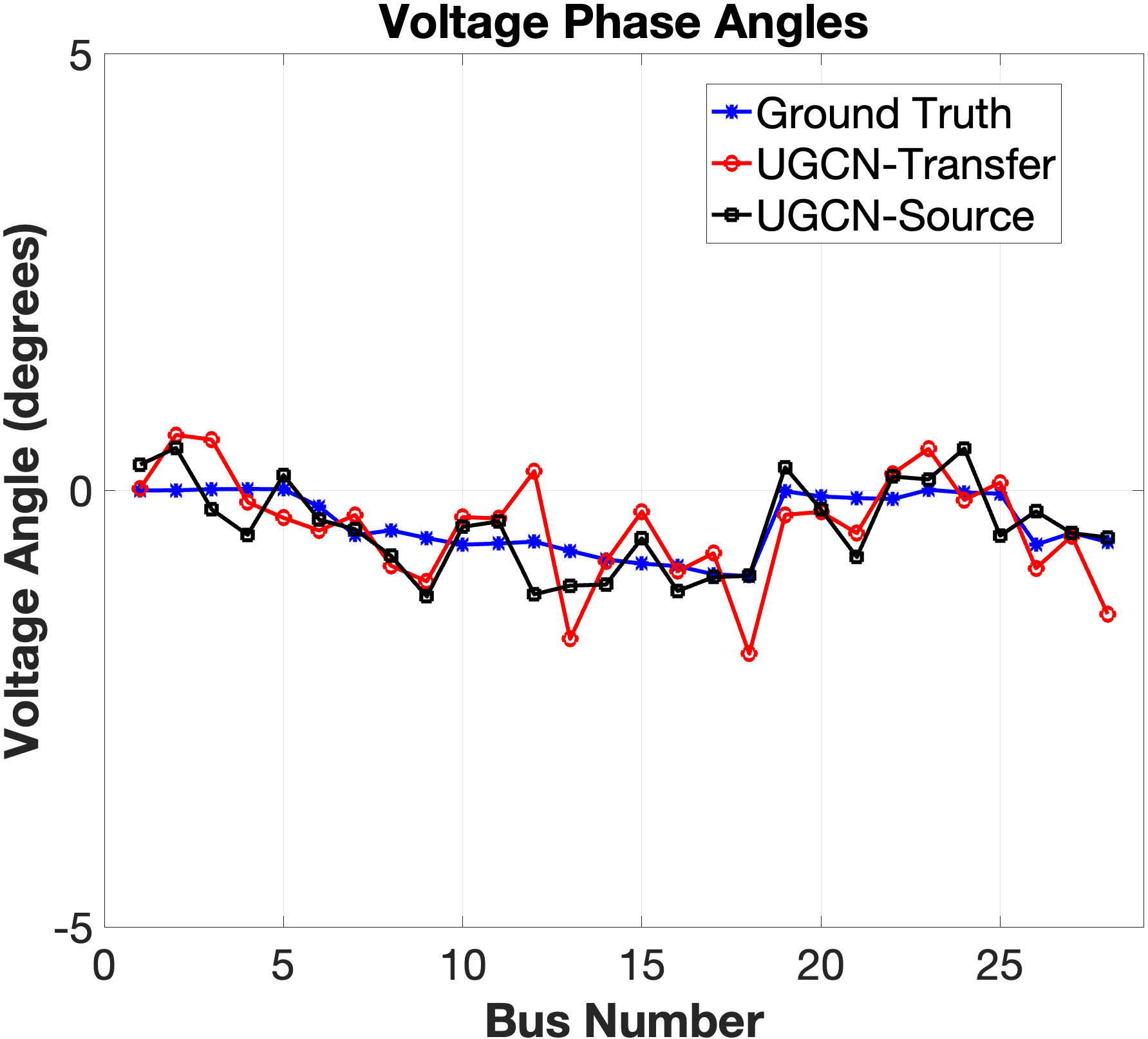}
     \label{fig:state_est_phase}
} 
  \subfigure[One-hour forecasting: Voltage magnitudes ($H=1$)]{
 \includegraphics[width=1.6in]{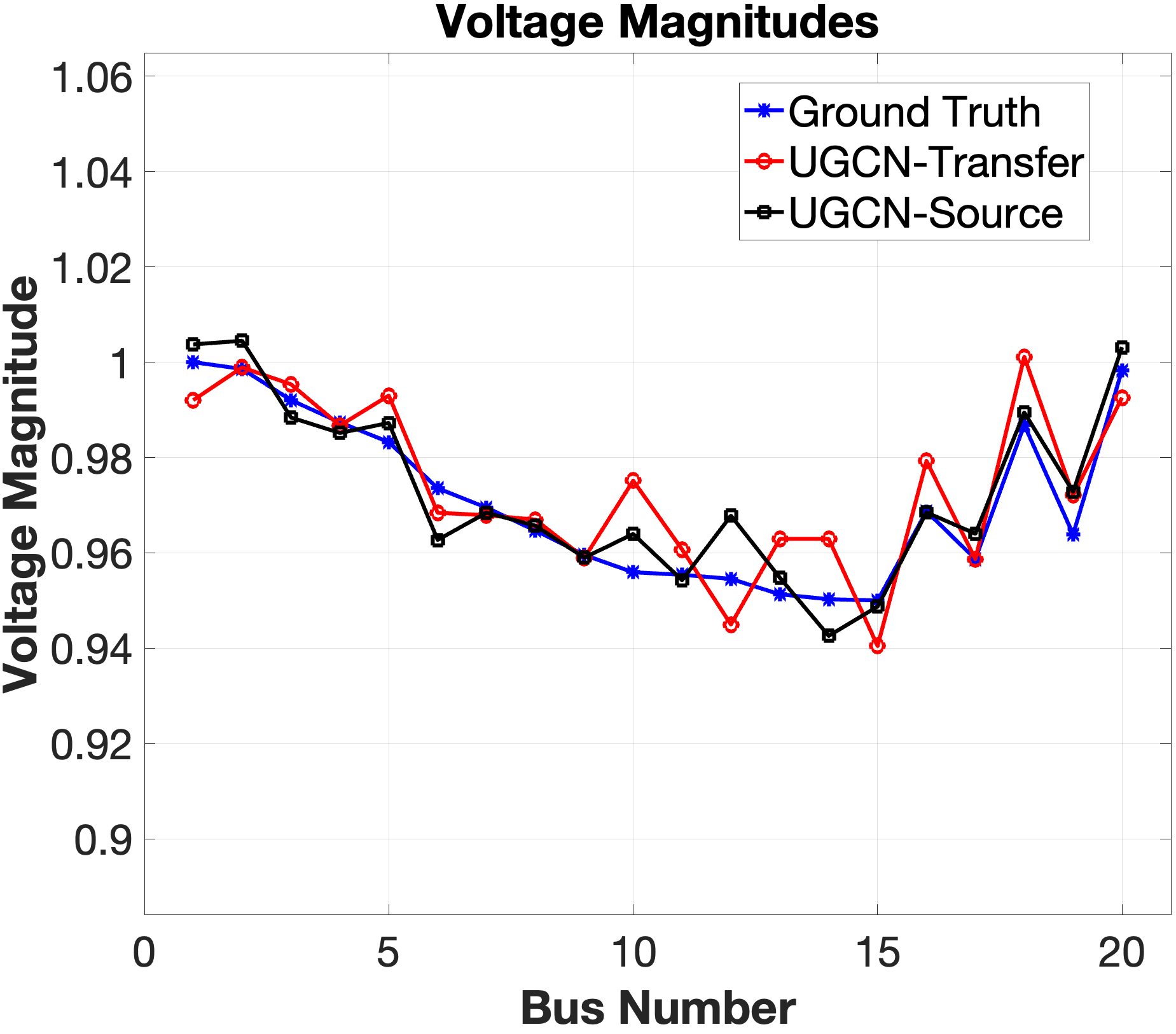}
      \label{fig:forecast_mag}
 } 
 \subfigure[One-hour forecasting: Voltage phase angles ($H=1$)]{
\includegraphics[width=1.6in]{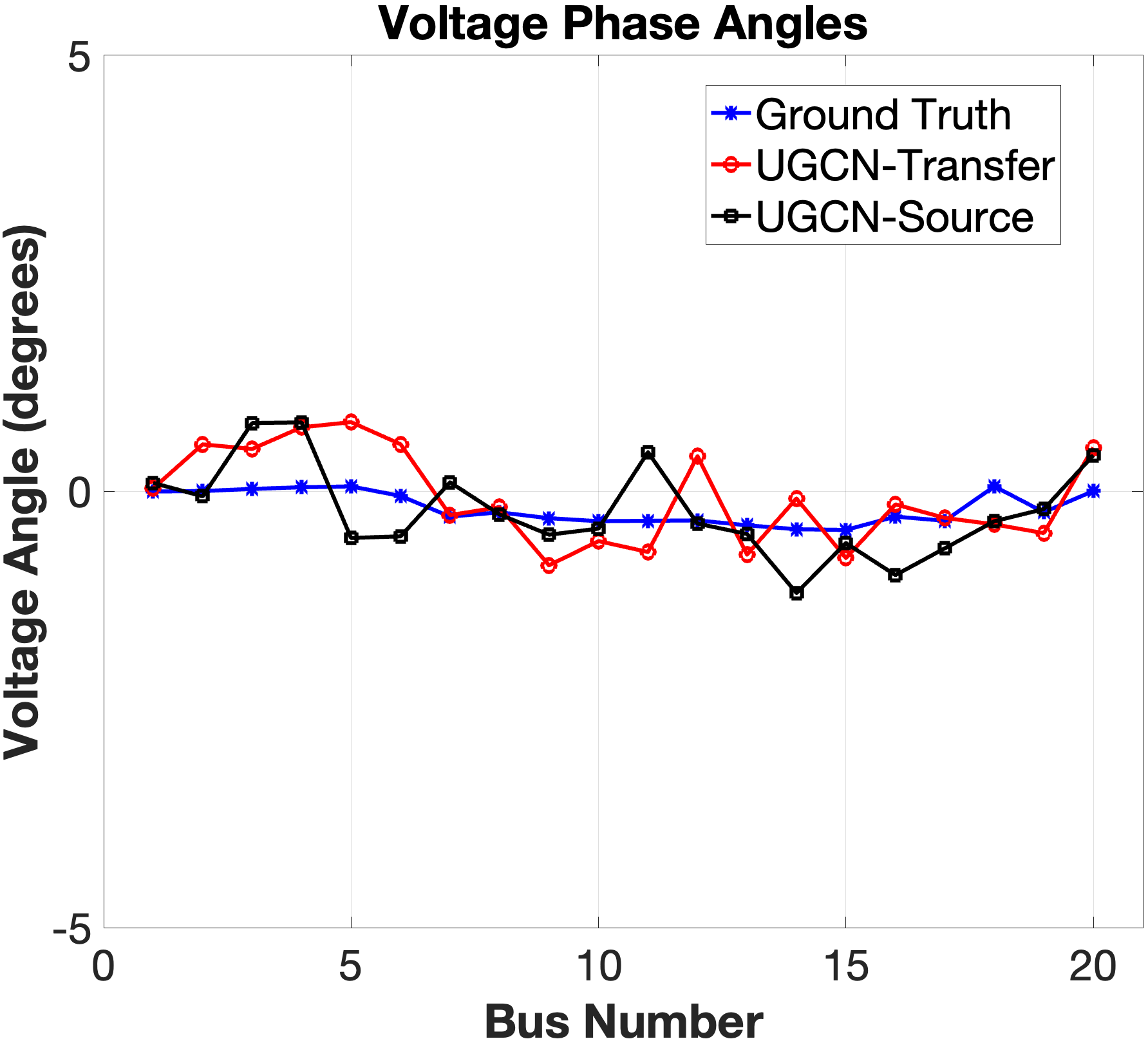}
     \label{fig:forecast_phase}
} 
  \caption{Zero-shot transfer example: UGCN state estimation and forecasting performance on unseen IEEE 69-bus system reconfiguration}
  \label{fig:estimation_forecast}
  \vspace{-0.6cm}
\end{figure}

\begin{figure*}[!htbp]
\centering
\begin{minipage}{1.00\textwidth}
    \centering
    \subfigure[Zero-shot transfer: FDI localization on unseen IEEE 30-bus reconfigurations]{\includegraphics[width=0.33\textwidth]{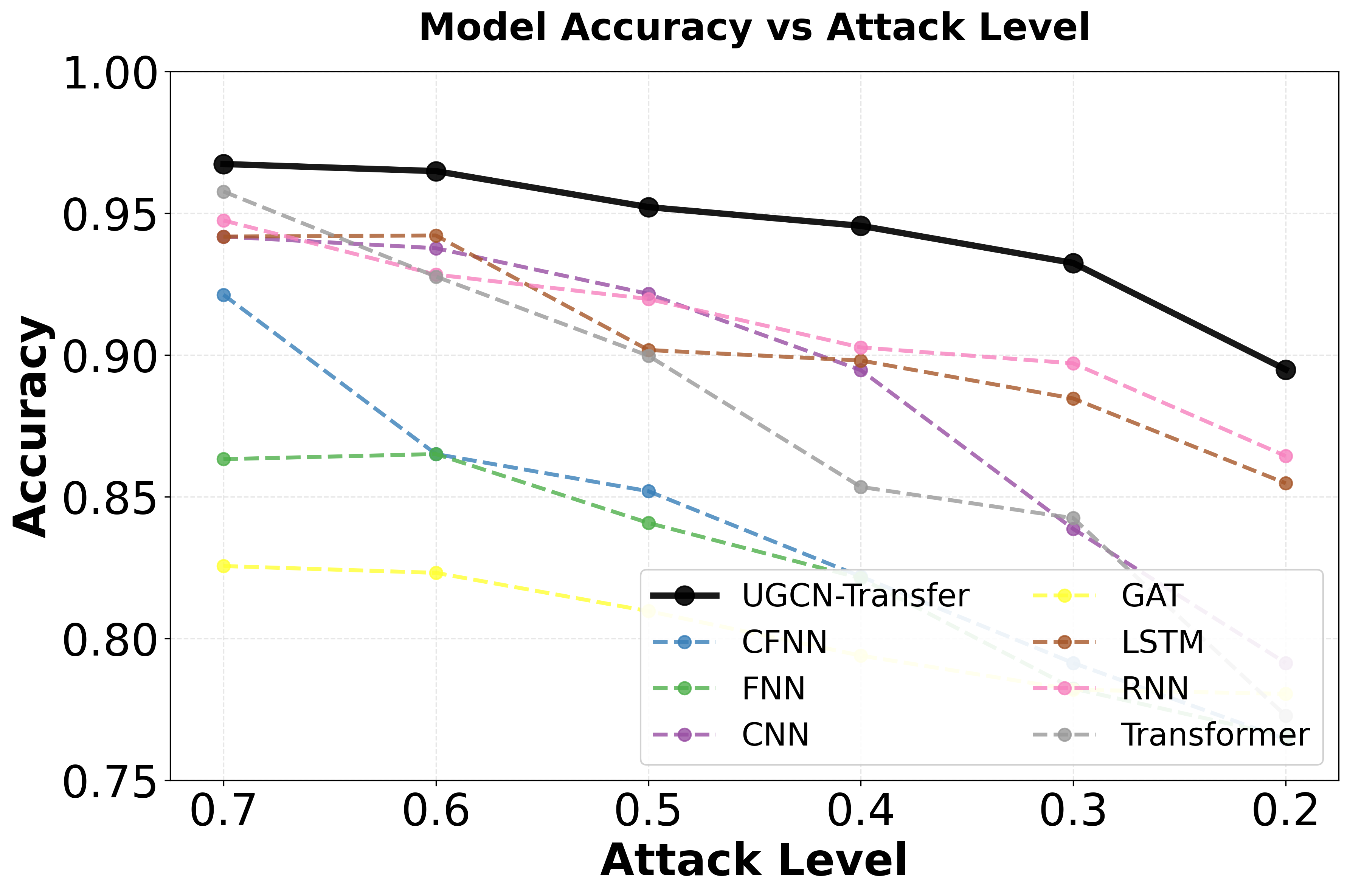}\label{fig:fdi_case30}}
    \hspace{-0.05in}
    \vspace{-0.05in}   
    \subfigure[Zero-shot transfer: FDI localization on unseen IEEE 39-bus reconfigurations]{\includegraphics[width=0.33\textwidth]{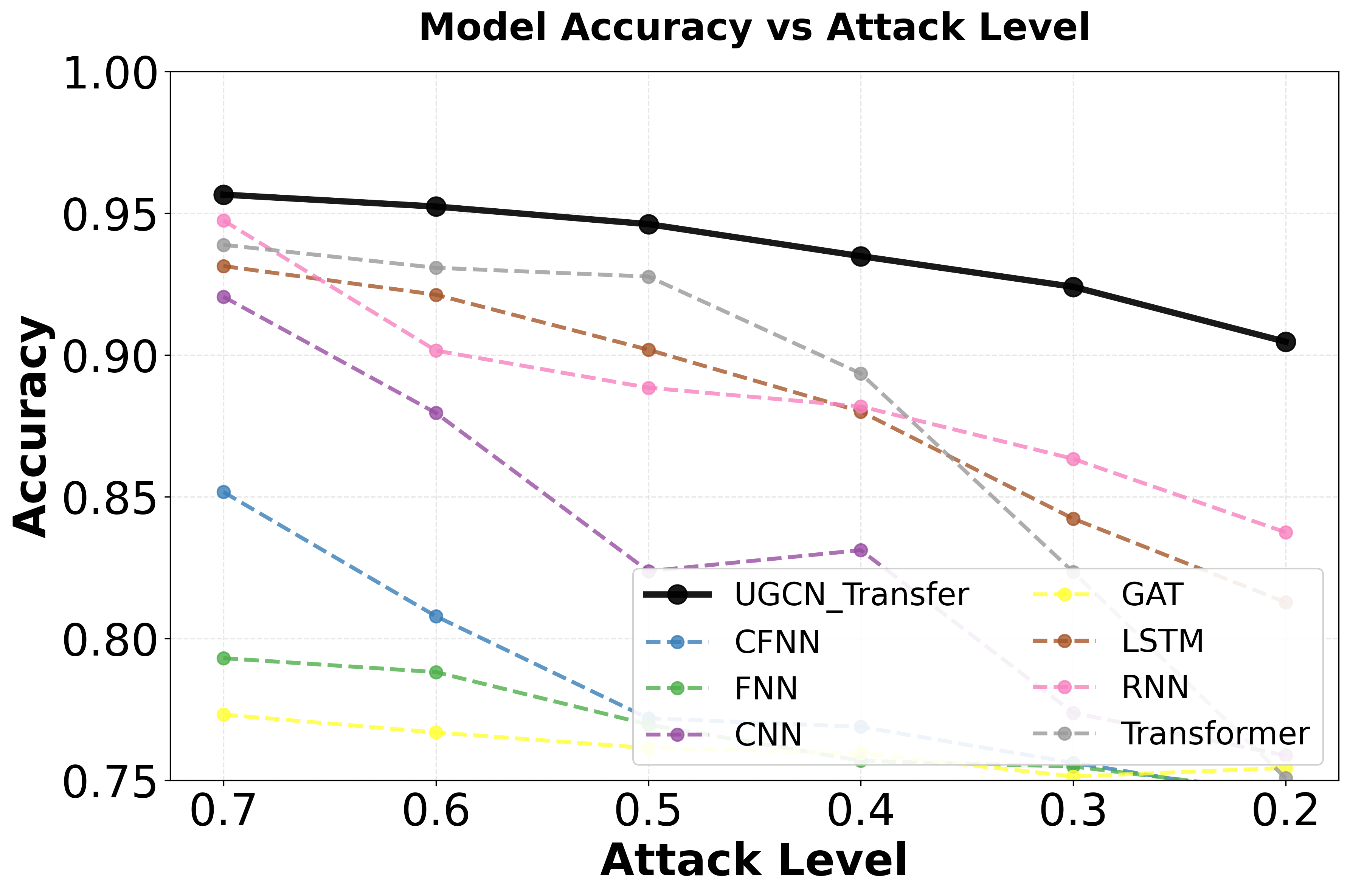}\label{fig:fdi_case39}}
    \hspace{-0.05in}
    \vspace{-0.05in}     
    \subfigure[Zero-shot transfer: FDI localization on unseen IEEE 57-bus reconfigurations]{\includegraphics[width=0.33\textwidth]{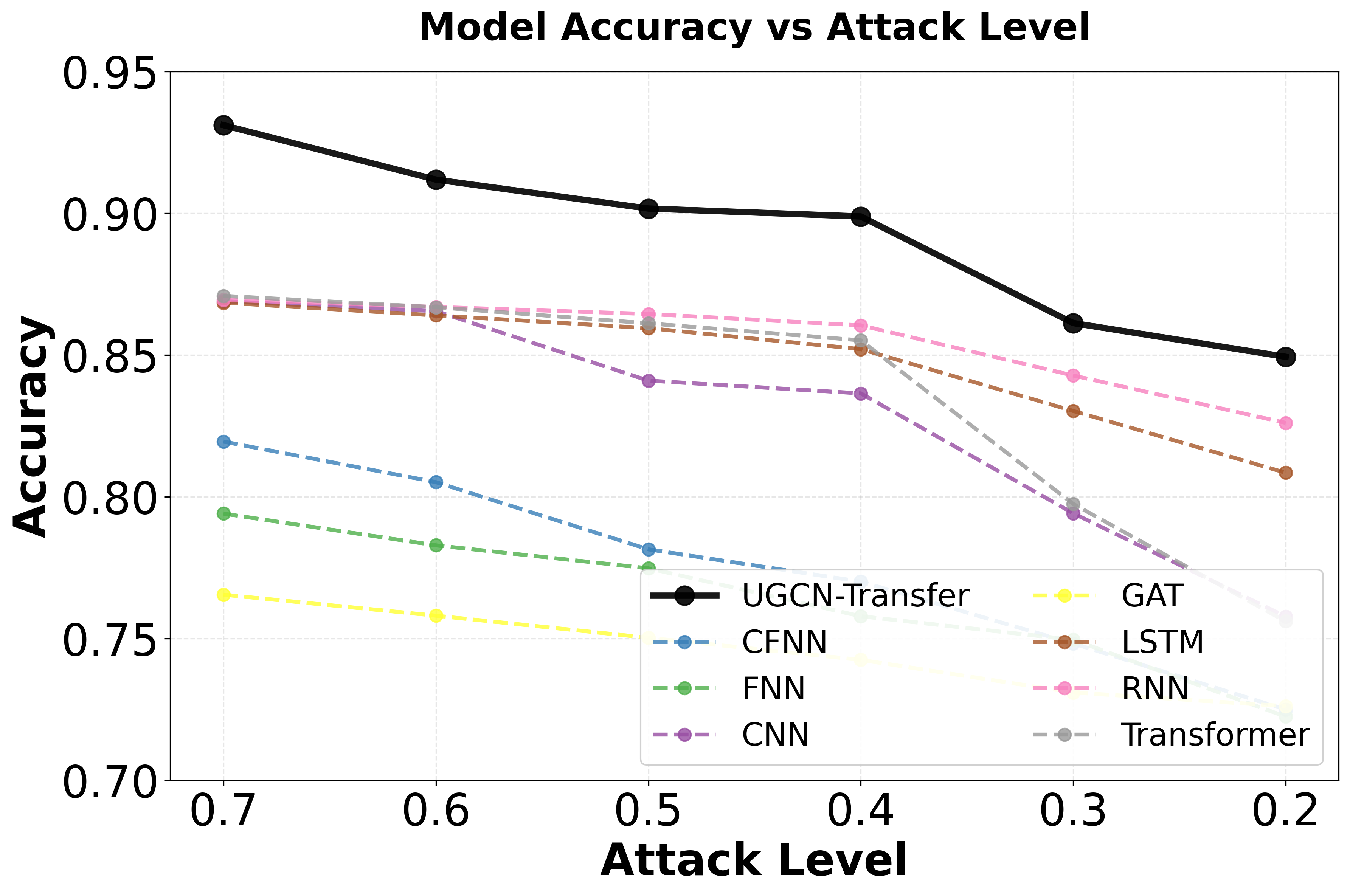}\label{fig:fdi_case57}}
    \caption{UGCN transferability for false data injection localization across different individual power system configurations without retraining.}
    \label{fig:fdi_individual}
\end{minipage}
\vspace{-0.6cm}
\end{figure*} 
\begin{figure}[!htbp]  
    \centering
    \subfigure[Heterogeneous Multi-Grid Transfer using one UGCN: FDI localization across IEEE 30 \& 39-bus unseen new reconfigurations]{\includegraphics[width=0.24\textwidth]{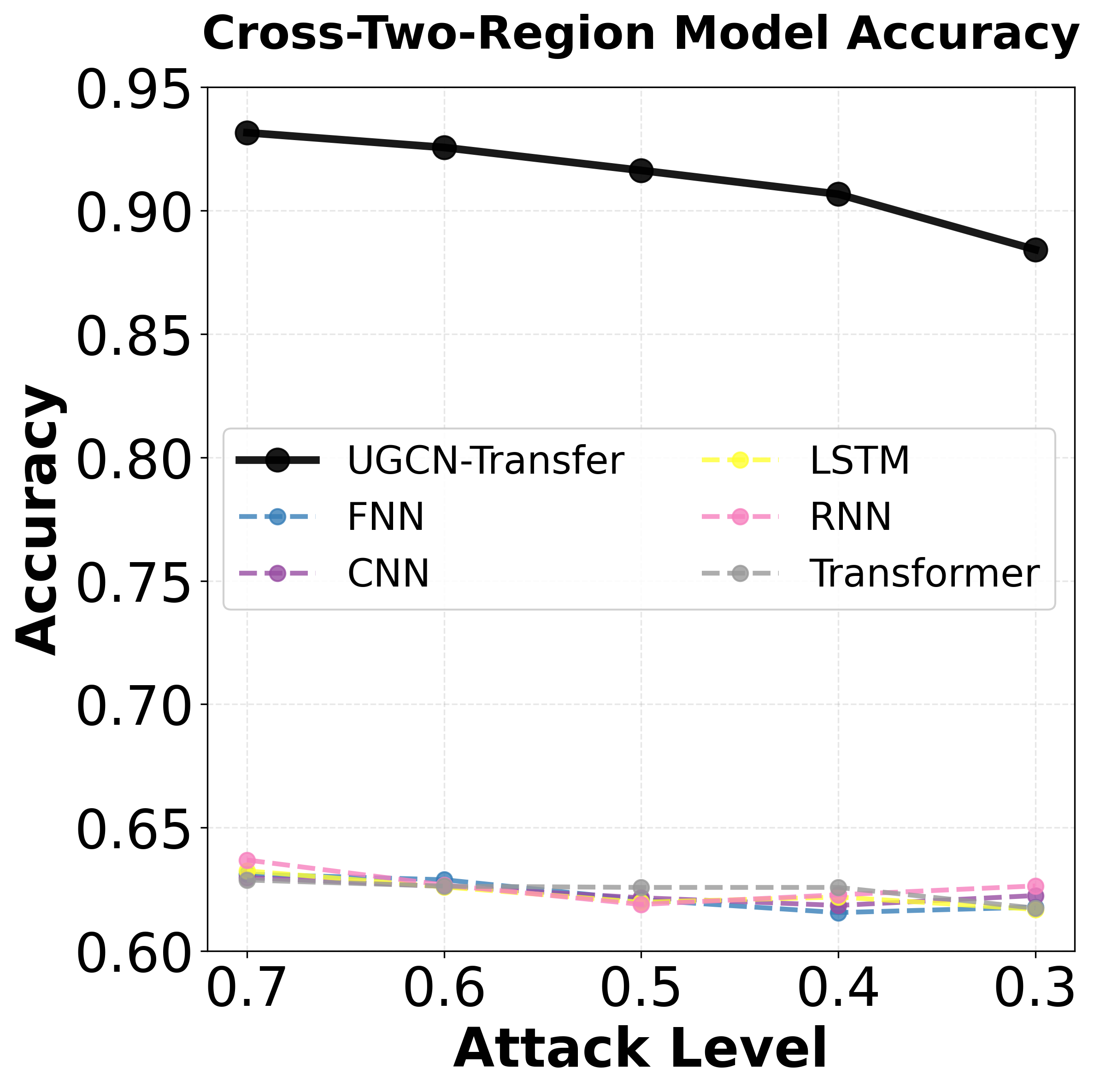}\label{fig:fdi_hybrid_30_39}}
    \hspace{-0.05in} 
    \vspace{-0.1in}  
    \subfigure[Heterogeneous Multi-Grid Transfer using one UGCN: FDI localization across IEEE 30 \& 39 \& 57-bus unseen new reconfigurations]{\includegraphics[width=0.24\textwidth]{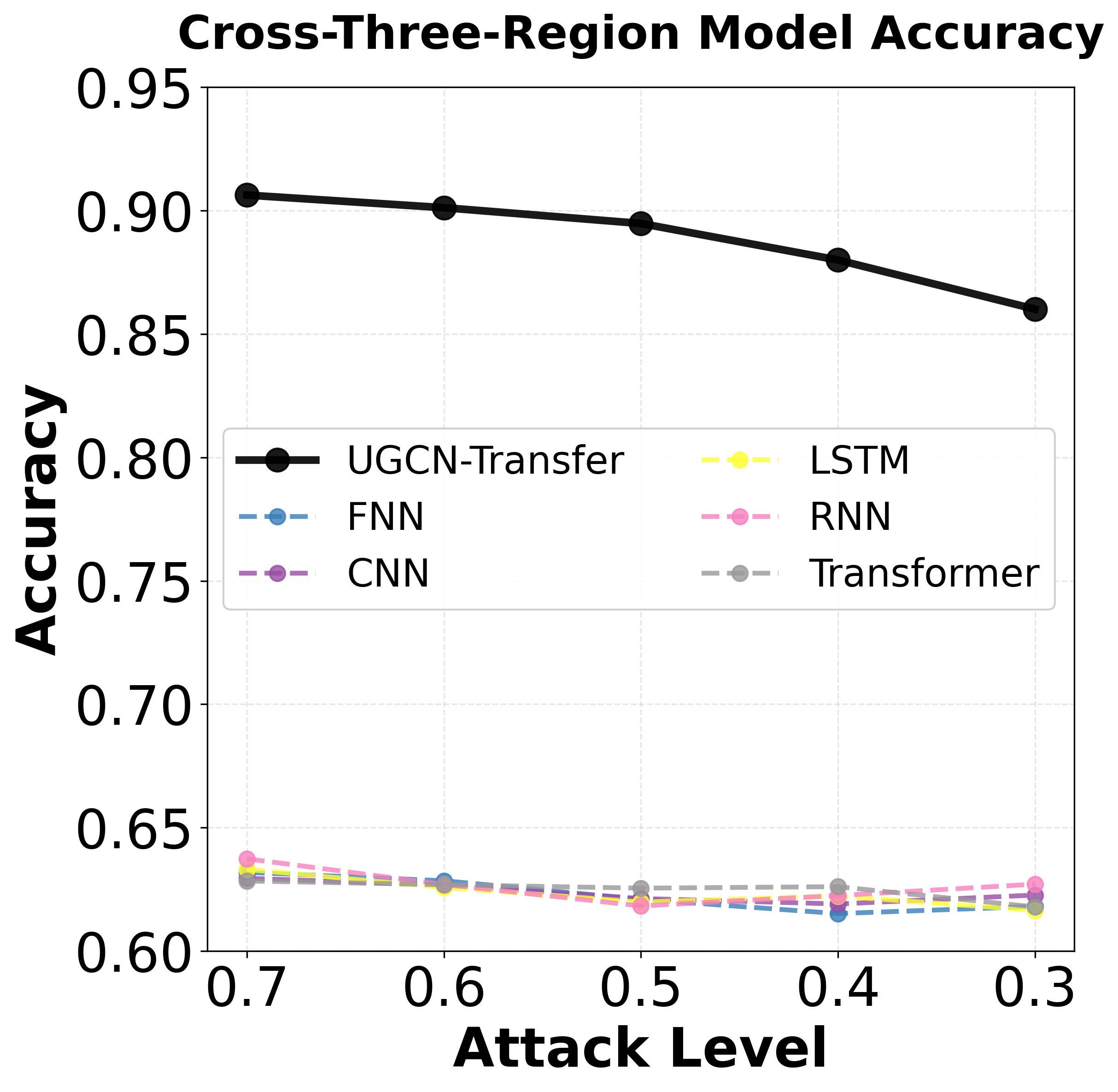}\label{fig:fdi_hybrid_30_39_57}}
    \caption{UGCN transferability for heterogeneous multi-grid FDI localization across multiple unseen power system reconfigurations}
    \label{fig:fdi_hybrid}
\vspace{-0.6cm} 
\end{figure}

\subsection{Power Distribution System State Forecasting}
\subsubsection{Experimental Setup}  
All experiments use a 10-hour historical window ($T=10$) to forecast voltage phasors for the next 1--5 hours ($H=1,2,3,4,5$). When $H=0$, the problem reduces to power system state estimation. We evaluate two measurement scenarios: \textbf{Scenario 1 (AMI)} uses AMI measurements on 40\% of buses with DistFlow-based optimization \eqref{eq:distflow_sse}, while \textbf{Scenario 2 (Sparse PMU)} uses microPMUs on 20\% of buses with voltage estimation via regularized least squares \eqref{eq:state_estimation}. Estimated voltage phasors serve as forecasting inputs for both scenarios.

\subsubsection{Results on IEEE 33-Bus System}
We evaluate performance under various reconfiguration scenarios comparing UGCN against baseline models (FNN, CplxFNN, GRU, Transformer, GAT, RNN, CNN, LSTM) and their padded extensions.

For \textbf{Scenario 1 (AMI)}, UGCN-Source achieves the lowest MSE of $9.621 \times 10^{-5}$ at $H=5$, significantly outperforming baselines with MSEs ranging from $2.004 \times 10^{-2}$ (Transformer) to $6.889 \times 10^{-2}$ (RNN). UGCN-Transfer also demonstrates excellent performance with MSE of $1.047 \times 10^{-4}$, representing approximately a 191× improvement over the best competing method, as shown in Figure~\ref{fig:ami_33}.

For \textbf{Scenario 2 (Sparse PMU)}, UGCN-Source maintains superior performance with MSE of $9.634 \times 10^{-5}$ at $H=5$, compared to baseline MSEs ranging from $2.898 \times 10^{-2}$ (Transformer) to $6.006 \times 10^{-2}$ (RNN). UGCN-Transfer achieves MSE of $1.221 \times 10^{-4}$, representing approximately a 237× improvement over the best traditional method, as illustrated in Figure~\ref{fig:pmu_33}.

Figure~\ref{fig:estimation_forecast} provides detailed visualization of voltage magnitude and phase angle predictions, showing the close resemblance between UGCN predictions and ground-truth values for both state estimation ($H=0$) and one-hour forecasting ($H=1$) on IEEE 33-bus system reconfigurations.

\subsubsection{Results on IEEE 69-Bus System}
Similar evaluation is conducted on the IEEE 69-bus system for both scenarios. For \textbf{Scenario 1 (AMI)}, UGCN-Source achieves the lowest MSE of $6.257 \times 10^{-5}$ at $H=5$, compared to baseline range of $1.684 \times 10^{-2}$ (Transformer) to $7.308 \times 10^{-2}$ (RNN). UGCN-Transfer maintains excellent performance with MSE of $8.673 \times 10^{-5}$, demonstrating consistent superiority across different system sizes, as shown in Figure~\ref{fig:ami_69}.

For \textbf{Scenario 2 (Sparse PMU)}, UGCN achieves robust performance with MSE of $6.833 \times 10^{-5}$ and UGCN-Self attains $6.482 \times 10^{-5}$ at $H=5$, significantly outperforming traditional methods with MSEs ranging from $1.757 \times 10^{-2}$ (Transformer) to $5.025 \times 10^{-2}$ (RNN). This represents approximately a 271× improvement over the best competing algorithm, as illustrated in Figure~\ref{fig:pmu_69}.

These results demonstrate that UGCN maintains superior forecasting accuracy across different measurement types, system sizes, and configurations, confirming strong scalability and generalization capabilities. The consistent orders-of-magnitude improvements highlight UGCN's ability to capture fundamental power system physics that remain invariant across different network topologies.

\subsection{FDI Detection and Localization}
\subsubsection{FDI Setting}
We consider a binary classification task where each sensor can be either attacked or not. The IEEE 30-bus system has 15 sensors ($|\mathcal{A}| = 15$), IEEE 39-bus has 20 sensors, and the IEEE 57-bus system has 25 sensors, resulting in $2^{15}$, $2^{20}$, and $2^{25}$ possible attack combinations, respectively. The number of attacked buses varies randomly from zero to all sensors. The UGCN output is thresholded at 0.5 for classification, though this threshold can be adjusted based on application requirements. To capture multi-scale features, we adopt hybrid pooling that combines average and max adaptive pooling. Attack detectability increases with $\omega$ in \eqref{eq:attack_model}, where larger $\omega$ values make attacks easier to detect.

\subsubsection{Results on Individual IEEE Systems}
We evaluate the proposed UGCN model against baseline models, including fully-connected and graph-based architectures. The FDI localization performance on individual IEEE systems is presented in Figure~\ref{fig:fdi_individual}. Across all systems, UGCN consistently outperforms all baselines. While several models, especially fully-connected and LSTM-based architectures, tend to predict trivial solutions (e.g., all-ones or all-zeros) depending on the number of attacks $|\mathcal{C}|$, UGCN effectively captures spatiotemporal dependencies in voltage phasor dynamics. 

For the IEEE 30-bus system, UGCN-Transfer achieves the highest accuracy of 96.74\% at attack level 0.7, significantly outperforming the second-best Transformer (95.77\%) and RNN (94.75\%). For the IEEE 39-bus system, UGCN-Transfer maintains superior performance with 95.66\% accuracy at attack level 0.7, compared to RNN (94.75\%) and LSTM (93.14\%). Similarly, for the IEEE 57-bus system, UGCN-Transfer achieves 93.11\% accuracy, outperforming Transformer (87.09\%) and RNN (86.95\%).

\subsubsection{Results on Multi-System Hybrid Training}
Figure~\ref{fig:fdi_hybrid} demonstrates UGCN's ability to be trained once across multiple disparate transmission networks and transfer to their reconfigurations within a single model. For the IEEE 30 \& 39-bus hybrid system, UGCN-Transfer achieves remarkable performance with 93.16\% accuracy at attack level 0.7, while all baseline methods cluster around 62-64\% accuracy, demonstrating UGCN's exceptional capability in cross-system learning.

For the more challenging IEEE 30 \& 39 \& 57-bus hybrid system, UGCN-Transfer maintains strong performance with 90.63\% accuracy at attack level 0.7, still substantially outperforming all baseline algorithms which remain around 62-63\% accuracy. This represents approximately 27+ percentage points improvement over the best competing method, highlighting UGCN's superior transferability across heterogeneous power system architectures.

Baseline algorithms utilize padding strategies to handle varying input and output dimensions when training on multiple systems, but fail to capture the underlying power system physics that enable effective cross-system generalization.


\section{Conclusions}
This work addressed a fundamental challenge in applying machine learning to power systems by introducing the novel UGCN, which achieved transferability across power system reconfigurations without prior knowledge of new grid topologies or retraining. Our physics-aware approach leveraged spatio-temporal graph convolutions to capture universal relationships that remained consistent across different network architectures.
The UGCN framework incorporated graph augmentation strategies, adaptive pooling mechanisms, and   Transformer output layers. Comprehensive evaluation across power system state forecasting and FDI detection demonstrated superior generalization capabilities compared to state-of-the-art methods. Results confirmed superior performance across diverse systems and reconfiguration scenarios, offering significant practical advantages for real-world applications. Future work will transfer to beyound reconfiguration, will transfer to a totally new grids without any retrain or prior knowledge of the new grids. Future work will extend beyond reconfigurations to achieve transfer to entirely new grids without any retraining or prior knowledge of the new grid topologies.

\begin{footnotesize}

\end{footnotesize}


\begin{thebibliography}{10}
\providecommand{\url}[1]{#1}
\csname url@samestyle\endcsname
\providecommand{\newblock}{\relax}
\providecommand{\bibinfo}[2]{#2}
\providecommand{\BIBentrySTDinterwordspacing}{\spaceskip=0pt\relax}
\providecommand{\BIBentryALTinterwordstretchfactor}{4}
\providecommand{\BIBentryALTinterwordspacing}{\spaceskip=\fontdimen2\font plus
\BIBentryALTinterwordstretchfactor\fontdimen3\font minus
  \fontdimen4\font\relax}
\providecommand{\BIBforeignlanguage}[2]{{%
\expandafter\ifx\csname l@#1\endcsname\relax
\typeout{** WARNING: IEEEtran.bst: No hyphenation pattern has been}%
\typeout{** loaded for the language `#1'. Using the pattern for}%
\typeout{** the default language instead.}%
\else
\language=\csname l@#1\endcsname
\fi
#2}}
\providecommand{\BIBdecl}{\relax}
\BIBdecl

\bibitem{zhu2015optimization}
J.~Zhu, \emph{Optimization of power system operation}.\hskip 1em plus 0.5em
  minus 0.4em\relax John Wiley \& Sons, 2015.

\bibitem{wu2023constrained}
T.~Wu, A.~Scaglione, and D.~Arnold, ``Constrained reinforcement learning for
  predictive control in real-time stochastic dynamic optimal power flow,''
  \emph{IEEE Trans. Power Syst.}, vol.~39, no.~3, pp. 5077--5090, 2023.

\bibitem{aguero2017modernizing}
J.~R. Aguero, E.~Takayesu, D.~Novosel, and R.~Masiello, ``Modernizing the grid:
  Challenges and opportunities for a sustainable future,'' \emph{IEEE Power
  Energy Mag.}, vol.~15, no.~3, pp. 74--83, 2017.

\bibitem{su2024review}
T.~Su, T.~Wu, J.~Zhao, A.~Scaglione, and L.~Xie, ``A review of safe
  reinforcement learning methods for modern power systems,'' \emph{arXiv
  preprint arXiv:2407.00304}, 2024.

\bibitem{wu2023complex}
T.~Wu, A.~Scaglione, and D.~Arnold, ``Complex-value spatio-temporal graph
  convolutional neural networks and its applications to electric power systems
  ai,'' \emph{IEEE Trans. Smart Grid}, 2023.

\bibitem{li2022transfer}
H.~Li, Z.~Ma, and Y.~Weng, ``A transfer learning framework for power system
  event identification,'' \emph{IEEE Trans. Power Syst.}, vol.~37, no.~6, pp.
  4424--4435, 2022.

\bibitem{wu2024transferable}
T.~Wu, A.~Scaglione, D.~Arnold, and T.~Chen, ``Transferable learning of gcn
  sampling graph data clusters from different power systems,'' in \emph{2024
  60th Annual Allerton Conference on Communication, Control, and
  Computing}.\hskip 1em plus 0.5em minus 0.4em\relax IEEE, 2024, pp. 1--7.

\bibitem{khodayar2020deep}
M.~Khodayar, G.~Liu, J.~Wang, and M.~E. Khodayar, ``Deep learning in power
  systems research: A review,'' \emph{CSEE J. Power Energy Syst.}, vol.~7,
  no.~2, pp. 209--220, 2020.

\bibitem{hijazi2023transfer}
M.~Hijazi, P.~Dehghanian, and S.~Wang, ``Transfer learning for transient
  stability predictions in modern power systems under enduring topological
  changes,'' \emph{IEEE Trans. Autom. Sci. Eng.}, 2023.

\bibitem{niu2021decade}
S.~Niu, Y.~Liu, J.~Wang, and H.~Song, ``A decade survey of transfer learning
  (2010--2020),'' \emph{IEEE Trans. Artif. Intell.}, vol.~1, no.~2, pp.
  151--166, 2021.

\bibitem{wu2020voltage}
T.~Wu, Y.-J.~A. Zhang, and H.~Wen, ``Voltage stability monitoring based on
  disagreement-based deep learning in a time-varying environment,'' \emph{IEEE
  Trans. Power Syst.}, vol.~36, no.~1, pp. 28--38, 2020.

\bibitem{shi2022power}
J.~Shi, K.~Yamashita, and N.~Yu, ``Power system event identification with
  transfer learning using large-scale real-world synchrophasor data in the
  united states,'' in \emph{2022 IEEE Power \& Energy Society Innovative Smart
  Grid Technologies Conference (ISGT)}.\hskip 1em plus 0.5em minus 0.4em\relax
  IEEE, 2022, pp. 1--5.

\bibitem{li2022adaptive}
B.~Li and J.~Wu, ``Adaptive assessment of power system transient stability
  based on active transfer learning with deep belief network,'' \emph{IEEE
  Trans. Autom. Sci. Eng.}, vol.~20, no.~2, pp. 1047--1058, 2022.

\bibitem{ren2021integrated}
C.~Ren, Y.~Xu, B.~Dai, and R.~Zhang, ``An integrated transfer learning method
  for power system dynamic security assessment of unlearned faults with missing
  data,'' \emph{IEEE Trans. Power Syst.}, vol.~36, no.~5, pp. 4856--4859, 2021.

\bibitem{zheng2024meta}
L.~Zheng, Y.~Zhu, and Y.~Zhou, ``Meta-transfer learning-based method for
  multi-fault analysis and assessment in power system,'' \emph{Appl. Intell.},
  vol.~54, no.~23, pp. 12\,112--12\,127, 2024.

\bibitem{sun2019meta}
Q.~Sun, Y.~Liu, T.-S. Chua, and B.~Schiele, ``Meta-transfer learning for
  few-shot learning,'' in \emph{Proceedings of the IEEE/CVF conference on
  computer vision and pattern recognition}, 2019, pp. 403--412.

\bibitem{xia2024efficient}
W.~Xia, G.~Peng, C.~Wang, P.~Palensky, E.~Pauwels, and P.~P. Vergara, ``An
  efficient and explainable transformer-based few-shot learning for modeling
  electricity consumption profiles across thousands of domains,'' \emph{arXiv
  preprint arXiv:2408.08399}, 2024.

\bibitem{kim2023transient}
J.~Kim, H.~Lee, S.~Kim, S.-H. Chung, and J.~H. Park, ``Transient stability
  assessment using deep transfer learning,'' \emph{IEEE Access}, vol.~11, pp.
  116\,622--116\,637, 2023.

\bibitem{shi2022bidirectional}
Z.~Shi, W.~Yao, Y.~Tang, X.~Ai, J.~Wen, and S.~Cheng, ``Bidirectional active
  transfer learning for adaptive power system stability assessment and dominant
  instability mode identification,'' \emph{IEEE Trans. Power Syst.}, vol.~38,
  no.~6, pp. 5128--5142, 2022.

\bibitem{liang2024unified}
H.~Liang, C.~Zhao, and M.~Chen, ``A unified deep neural network for solving ac
  opf in expanding and multiple networks,'' in \emph{2024 IEEE International
  Conference on Communications, Control, and Computing Technologies for Smart
  Grids (SmartGridComm)}.\hskip 1em plus 0.5em minus 0.4em\relax IEEE, 2024,
  pp. 575--580.

\bibitem{wu2023spatio}
T.~Wu, I.~L. Carre{\~n}o, A.~Scaglione, and D.~Arnold, ``Spatio-temporal graph
  convolutional neural networks for physics-aware grid learning algorithms,''
  \emph{IEEE Trans. Smart Grid}, 2023.

\bibitem{meilua2024manifold}
M.~Meil{\u{a}} and H.~Zhang, ``Manifold learning: What, how, and why,''
  \emph{Annual Review of Statistics and Its Application}, vol.~11, no.~1, pp.
  393--417, 2024.

\bibitem{krishnan2018challenges}
R.~Krishnan, D.~Liang, and M.~Hoffman, ``On the challenges of learning with
  inference networks on sparse, high-dimensional data,'' in \emph{International
  conference on artificial intelligence and statistics}.\hskip 1em plus 0.5em
  minus 0.4em\relax PMLR, 2018, pp. 143--151.

\bibitem{kipf2017semi}
T.~N. Kipf and M.~Welling, ``Semi-supervised classification with graph
  convolutional networks,'' in \emph{ICLR}, 2017.

\bibitem{vaswani2017attention}
A.~Vaswani, N.~Shazeer, N.~Parmar, J.~Uszkoreit, L.~Jones, A.~N. Gomez,
  {\L}.~Kaiser, and I.~Polosukhin, ``Attention is all you need,'' \emph{Adv.
  Neural Inf. Process. Syst.}, vol.~30, 2017.

\bibitem{velickovic2017graph}
P.~Velickovic, G.~Cucurull, A.~Casanova, A.~Romero, P.~Lio, Y.~Bengio
  \emph{et~al.}, ``Graph attention networks,'' \emph{stat}, vol. 1050, no.~20,
  pp. 10--48\,550, 2017.

\bibitem{rampasek2022GPS}
L.~Ramp\'{a}\v{s}ek, M.~Galkin, V.~P. Dwivedi, A.~T. Luu, G.~Wolf, and
  D.~Beaini, ``{Recipe for a General, Powerful, Scalable Graph Transformer},''
  \emph{Adv. Neural Inf. Process. Syst.}, vol.~35, 2022.

\bibitem{zhang2019power}
L.~Zhang, G.~Wang, and G.~B. Giannakis, ``Power system state forecasting via
  deep recurrent neural networks,'' in \emph{IEEE ICASSP}, 2019.

\bibitem{wang2020locational}
S.~Wang, S.~Bi, and Y.-J.~A. Zhang, ``Locational detection of the false data
  injection attack in a smart grid: A multilabel classification approach,''
  \emph{IEEE Internet Things J.}, vol.~7, no.~9, pp. 8218--8227, 2020.

\bibitem{wang2021kfrnn}
Y.~Wang, Z.~Zhang, J.~Ma, and Q.~Jin, ``Kfrnn: an effective false data
  injection attack detection in smart grid based on kalman filter and recurrent
  neural network,'' \emph{IEEE Internet Things J.}, vol.~9, 2021.

\end{thebibliography}
\end{document}